\documentclass[%
 reprint,
 amsmath,amssymb,
 aps,
]{revtex4-2}

\usepackage{CJKutf8}
\usepackage{array} 
\usepackage{multirow}
\usepackage{graphicx}
\usepackage{dcolumn}
\usepackage{bm}
\usepackage{xcolor}
\usepackage{ulem}

\begin{document}

\begin{CJK*}{GB}{} 

\title{Destinies of Plasmoids Formed from the Magnetic Reconnection Occurring within the Ergo Region of a Rotating Black Hole}

\author{Ho-Yun YuChih$^{2}$}
\author{\begin{CJK}{UTF8}{gbsn}
Ye Shen (沈冶)
\end{CJK}$^{1}$}%
\email{shenye199594@fzu.edu.cn}
\affiliation{
$~^{1}$ College of Physics and Information Engineering, Fuzhou University, Fuzhou 350108, Fujian, P. R. China \\
$~^{2}$ School of Physics, Peking University, No.5 Yiheyuan Rd, Beijing
		100871, P.R. China 
}%

\date{\today}

\begin{abstract}

It is widely accepted that magnetic reconnections occur successively in black hole accretion flows, especially on the equatorial plane, owing to the formation of the equatorial current sheet. When magnetic reconnection takes place within the ergosphere of a rotating black hole, one of the ejected plasmoids (denoted as p$^-$) may acquire negative energy-at-infinity density, while the companion plasmoid (denoted as p$^+$) correspondingly carries the energy extracted from the black hole. This mechanism, known as the Comisso-Asenjo process, provides a viable astrophysical pathway for extracting energy from a rotating black hole. Generally speaking, two conditions have to be satisfied for successful energy extraction. First, p$^-$ is generated via magnetic reconnection with negative energy-at-infinity density. Second, the associated p$^+$, which carries the extracted energy, must escape to infinity. Though a number of studies have promoted the theory of the Comisso-Asenjo process, the escape of p$^+$ lacks consideration. In this work, we adopt geodesic approach to systematically examine the destinies of p$^+$ after ejection, thereby establishing the complete conditions for energy extraction from a rotating black hole via magnetic reconnection. We redefine the covering factor, for which only escaping p$^+$ are counted. This quantity reflects how probable the successful energy extraction occurs in an accretion system. Through analysis of this revised covering factor, we identify a lower bound on the black hole spin, $a=0.8005$, below which energy extraction via magnetic reconnection can never succeed, no matter how high the magnetization of the bulk plasma is. This limit suggests that the spin-down of a black hole caused by successive magnetic reconnections would terminate once the spin decreases to approximately 0.8005.

\end{abstract}


\maketitle

\end{CJK*}


\section{Introduction}
\label{sec:intro}

It is widely accepted that plasma flows around supermassive black holes under the influence of strong magnetic field, within which the magnetic reconnection occurs frequently. As an efficient mechanism for converting electromagnetic energy into the kinetic or thermal energy of plasma, magnetic reconnection plays a critical role in interpreting a broad range of astrophysical phenomena observed in black hole accretion systems, such as the formation of hot spots \cite{Ripperda:2019lsi,Ripperda:2020bpz,Dexter:2020cuv,Lin:2023tog,Aimar2023,Nathanail:2024efu,Chen:2024jkm,Zeng:2026ufr} and episodic flaring in relativistic jets \cite{Cemeljic:2022rlc,Xi:2024sxb,Jiang:2025huk}. 

Within the framework of force-free electrodynamics \cite{Komissarov:2004ms,McKinney:2006sc,Gralla:2014yja,GS-Shen}, the so-called split-monopole magnetic configuration has been proposed, which predicts the emergence of antiparallel magnetic fields near the equatorial plane of a rotating black hole \cite{Split-Mono,Camilloni:2022kmx,Gelles:2024tpz,Song:2025mhj}. This magnetic structure, referred to as the equatorial current sheet, has already been supported by numerical simulations \cite{Davis:2020wea,Vos:2024loa,Yuan:2025war,Xia:2026swl}. The formation of such antiparallel structure leads to recurrent magnetic reconnection events on the equatorial plane, driven by the tearing mode instability in magnetofluids \cite{tearing-mode,Ebrahimi:2016urr,Comisso:2016pyg,Comisso:2017arh}. The occurrences of magnetic reconnection near a rotating black hole, particularly those within the ergosphere, render the so-called Comisso-Asenjo process feasible.

It is well known that magnetic reconnection gives rise to a pair of plasmoids. When it occurs within the ergo region of a rotating black hole, one of the plasmoids may carry energy extracted from the central black hole. This mechanism was initially considered by Koide and Arai \cite{KA2008} and subsequently developed by Comisso and Asenjo \cite{CA2021}. It offers an astrophysically plausible route for energy extraction from rotating black holes. This so-called Comisso-Asenjo process has attracted considerable attention in recent years, and has been verified through numerical simulations \cite{Xia:2026nkt}.

Numerous subsequent studies have further investigated the Comisso-Asenjo process. In Ref.~\cite{work0,work1,work2}, the scenario involving plunging bulk plasma was carefully examined. Additionally, the covering factor was defined preliminarily to quantify the capability of a black hole accretion system to extract energy from its central black hole via magnetic reconnection. By comparing the covering factor, it was revealed that the plunging bulk plasma is more conducive to energy extraction than the circularly flowing one. In Ref.~\cite{Camilloni:2024tny}, the occurrences of the Comisso-Asenjo process on the surface of a magnetized torus around a rotating black hole were explored. Beyond these foundational investigations, the feasibility of energy extraction from rotating black holes via magnetic reconnection has also been extended to a variety of spacetimes \cite{Carleo:2022qlv,Khodadi:2022dff,Wei:2022jbi,Liu:2022qnr,Wang:2022qmg,Shaymatov:2023dtt,ZhangShaoJun2024,Long:2024tws,Zeng:2025vjt,Zeng:2025olq,work3,Suleiman:2026iul,Yao:2026cvs,Zhao:2025uuk}.

Although efforts have continued to refine the theoretical understanding of the Comisso-Asenjo process, insufficient attention has been paid to whether the produced plasmoids can actually escape the gravitational trap of the central black hole. In previous works, energy extraction is considered successful as long as one of the plasmoids (denoted as p$^-$ for brevity) carries negative energy-at-infinity density, which equivalently implies that the other one (denoted as p$^+$) acquires the energy extracted from the central black hole. However, as one might expect, successful energy extraction would never be achieved if p$^+$ ultimately falls into the black hole with the extracted energy returned subsequently. Ref.~\cite{work0} briefly touched upon this issue, yet only within a restricted scenario. In this work, we aim to investigate this problem in a more comprehensive and systematic manner. In actual, there are a variety of methods for p$^+$ to send the extracted energy outward, such as radiation, advective transport and thermal conduction. However, it is nontrivial to address all these processes. We just focus on the direct escape of p$^+$ in this work.

Given that reconnection is thought to frequently occur in regions where the bulk plasma is sparse \cite{Chernoglazov:2023ksi,Oikawa:2026gvr,Figueiredo:2026qbx}, the generated plasmoids are likely to move with negligible influence from the surrounding plasma. Moreover, magnetic field is unlikely to affect their motions significantly because the plasmoids are nearly neutral and demagnetized. Thus, throughout this work, we treat the trajectories of plasmoids as geodesics after their formations. We calculate their energy-at-infinity density and angular-momentum-at-infinity density, which are the two conserved quantities for equatorial geodesics. By classifying the different types of geodesics, together with the conserved quantities, we determine the destinies of the plasmoids.

The destinies of p$^+$ generated with varying orientation angle $\xi_{B}$ at different radial positions of the reconnection point $r_0$ are presented in $\xi_B-r_0$ parameter planes for multiple values of black hole spin and bulk plasma magnetization. Additionally, we compare the covering factors, which quantify the capability of an accretion system to extract energy from the central black hole via magnetic reconnection, with and without accounting for the destinies of p$^+$. We will show that the likelihood of successful energy extraction decreases considerably for relatively low black hole spin ($a\lesssim 0.9$), as a large fraction of p$^+$ plunges into the black hole. However, when the spin of the central black hole becomes extreme ($a\rightarrow 1$), most generated plasmoids carrying extracted energy escape. The covering factors, with and without considering the destinies of plasmoids, are found to be close. More importantly, we find that successful energy extraction via magnetic reconnection could never occur if the spin of central black hole is below $a=0.8005$, regardless of the bulk plasma magnetization. This establishes a theoretical lower bound on the black hole spin required for viable energy extraction via this mechanism.

The remainder of this paper is organized as follows. We revisit some basic concepts of Comisso-Asenjo process in Sect.~\ref{sec:revisit}. We show how we judge the destinies of different plasmoids in Sect.~\ref{sec:geodesic}. Our main results are exhibited in Sect.~\ref{sec:results}, including the analyses on parameter planes shown in Sect.~\ref{sec:para-planes} and the variations of covering factors shown in Sect.~\ref{sec:covering}. We make a brief summary and further discussions in Sect.~\ref{sec:sum}.

\section{Comisso-Asenjo Process}
\label{sec:revisit}

We begin by revisiting some fundamental aspects of the Comisso-Asenjo process. The following analysis is carried out in the Kerr spacetime, whose line element can be conveniently expressed in the 3+1 form as:
\begin{equation}
    ds^2=g_{\mu\nu}dx^{\mu}dx^{\nu}=-\alpha^2dt^2+\sum_{i=1}^3\left(h_idx^i-\alpha\beta^idt\right)^2
    \label{eq:line}
\end{equation}
where $\alpha$ is the lapse function, $\beta^i=h_i\omega^i/\alpha$ is the shift vector with $\omega^i$ being the velocity of frame dragging and $h_i$ being the scale factor. 

Since reconnection takes place on the equatorial current sheet, the problem can be reduced to two dimensions. The tetrad of fluid's rest frame flowing on the equatorial plane is given by:
\begin{equation}
    \begin{aligned}
        e_{[0]}^{\mu}&=\hat{\gamma}_s \left[\hat{e}_{(t)}^{\mu}+\hat{v}_s^{(r)}\hat{e}_{(r)}^{\mu}+\hat{v}_s^{(\phi)}\hat{e}_{(\phi)}^{\mu}\right], \\
        e_{[1]}^{\mu}&=\frac{1}{\hat{v}_s}\left[\hat{v}_s^{(\phi)}\hat{e}_{(r)}^{\mu}-\hat{v}_s^{(r)}\hat{e}_{(\phi)}^{\mu}\right],~~
        e_{[2]}^{\mu}=\hat{e}_{(\theta)}^{\mu}, \\
        e_{[3]}^{\mu}&=\hat{\gamma}_s\left[\hat{v}_s\hat{e}_{(t)}^{\mu}+\frac{\hat{v}_s^{(r)}}{\hat{v}_s}\hat{e}_{(r)}^{\mu}+\frac{\hat{v}_s^{(\phi)}}{\hat{v}_s}\hat{e}_{(\phi)}^{\mu}\right]
    \end{aligned}
    \label{eq:tetrad-fluid}
\end{equation}
where $\hat{e}^{\mu}_{(\nu)}$ is the tetrad of zero-angular-momentum observers (ZAMOs):
\begin{equation}
    \begin{aligned}
        &~~~~~~~~~~~~~~
        \hat{e}_{(t)}^{\mu}=\frac{1}{\alpha}\left(\partial_t^{\mu}+\omega^{\phi}\partial_{\phi}^{\mu}\right); \\
        &\hat{e}_{(r)}^{\mu}=\frac{1}{h_r}\partial_r^{\mu},~~
        \hat{e}_{(\theta)}^{\mu}=\frac{1}{h_{\theta}}\partial_{\theta}^{\mu},~~
        \hat{e}_{(\phi)}^{\mu}=\frac{1}{h_{\phi}}\partial_{\phi}^{\mu}
    \end{aligned}
    \label{eq:tetrad-ZAMOs}
\end{equation}
In Eq.~\eqref{eq:tetrad-fluid}, $\hat{v}_s$ and $\hat{\gamma}_s$ are the speed and Lorentz factor of the bulk plasma, respectively, while $\hat{v}_s^{(r)}$ and $\hat{v}_s^{(\phi)}$ denote the radial and azimuthal components of 3-velocity, in the view of ZAMOs.

Two plasmoids (p$^+$ and p$^-$) are generated after magnetic reconnection. If reconnection takes place within the ergosphere, p$^+$ may carry energy extracted from the central black hole, which corresponds to the p$^-$ possessing negative energy-at-infinity density. The initial 4-velocities of the plasmoids immediately after ejection from the reconnection point are given by \cite{CA2021,work0}:
\begin{equation}
    u^{\mu}_{\rm p^{\pm},0}=\gamma_{\rm out}\left[e_{[0],0}^{\mu}\pm 
    v_{\rm out}\left(\cos\xi_{B}e_{[3],0}^{\mu}+\sin\xi_{B}e_{[1],0}^{\mu}\right)\right]
    \label{eq:u_out}
\end{equation}
where "$\pm$" corresponds to p$^+$ and p$^-$, respectively, and the subscript "0" denotes the reconnection point. Here, $\xi_B$ is the orientation angle, defined as the angle between the ejection direction of p$^+$ and the flowing direction of bulk plasma. $v_{\rm out}$ is the outflow speed of the plasmoids in the fluid's rest frame, namely, the speed at which the plasmoids are ejected from the reconnection point, satisfying \cite{work0}:
\begin{equation}
    v_{\rm out}\simeq \sqrt{\frac{\left(1-{\sf g}^2\right)^3\sigma_0}{\left(1+{\sf g}^2\right)^2+\left(1-{\sf g}^2\right)^3\sigma_0}}
    \label{eq:v_out}
\end{equation}
where $\sigma_0$ is the magnetization of the bulk plasma at the reconnection point and ${\sf g}$ is the geometric index relating to the microscopic magnetic structure \cite{Liu2017}. We adopt ${\sf g}\approx 0.49$, corresponding to the maximal reconnection rate in the high magnetization limit ($\sigma_0\gg 1$) \cite{work0}. In reality, the geometric index is an undetermined parameter. However, 0.49 is a relatively better choice as it corresponds to a higher reconnection rate. The influence of ${\sf g}$ in energy extraction is discussed in detail in Ref.~\cite{work2}.

\section{Geodesic Approach}
\label{sec:geodesic}

\begin{figure*}
    \centering
    \includegraphics[width=0.45\linewidth]{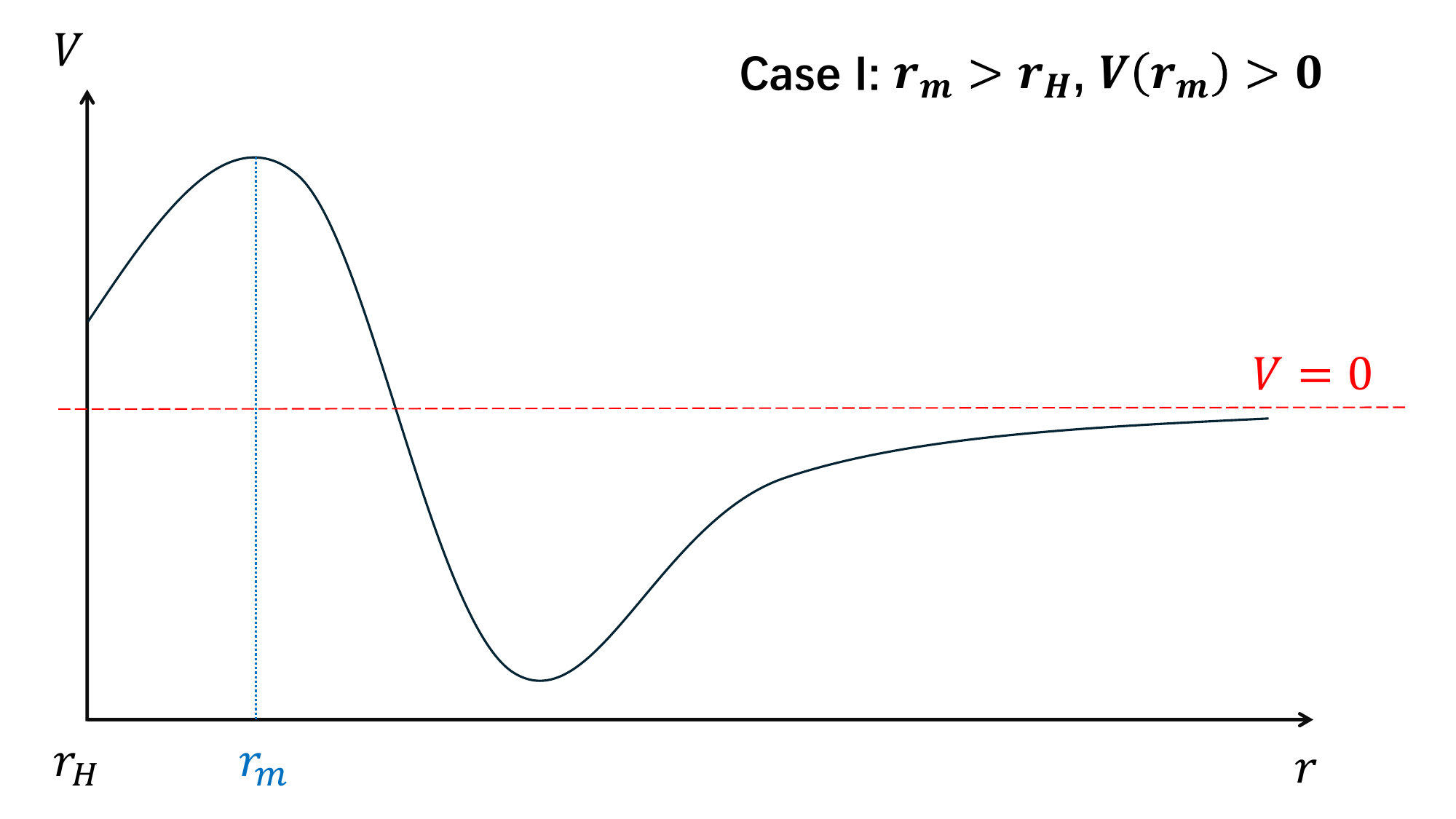}
    \includegraphics[width=0.45\linewidth]{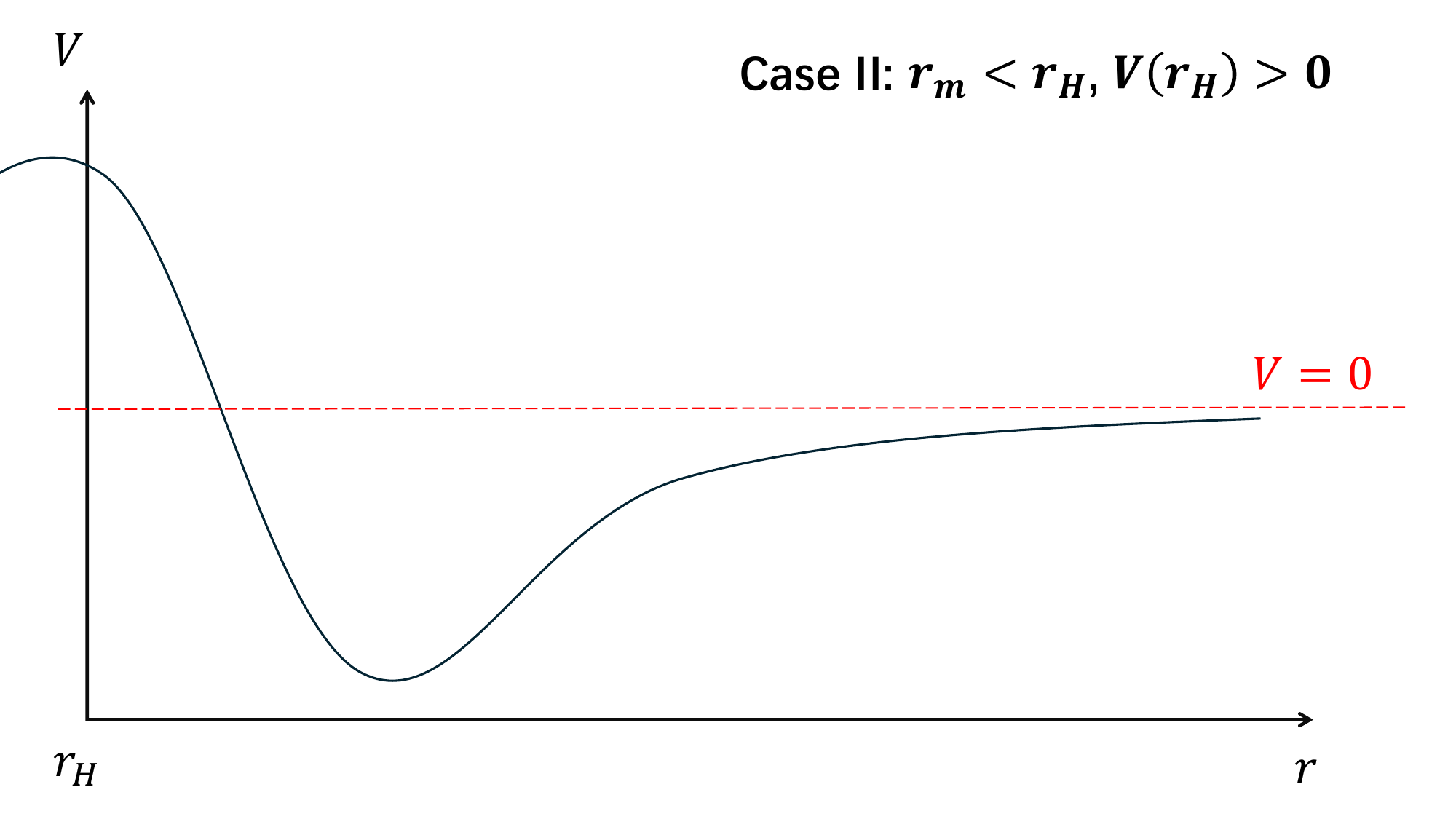}
    \\
    ~~~~~~~~~~~~~~~~~~~~~~
    \\
    \includegraphics[width=0.45\linewidth]{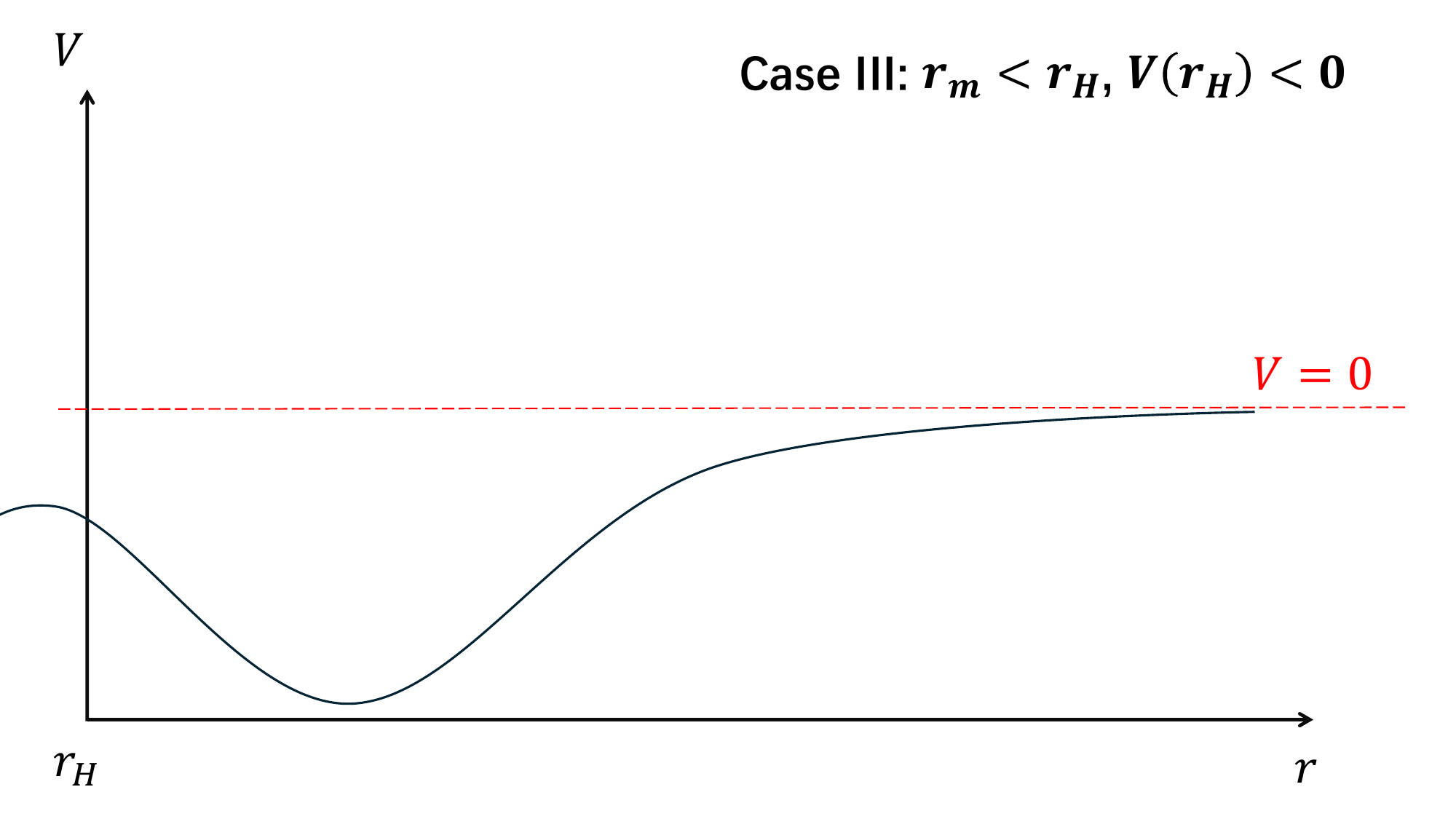}
    \includegraphics[width=0.45\linewidth]{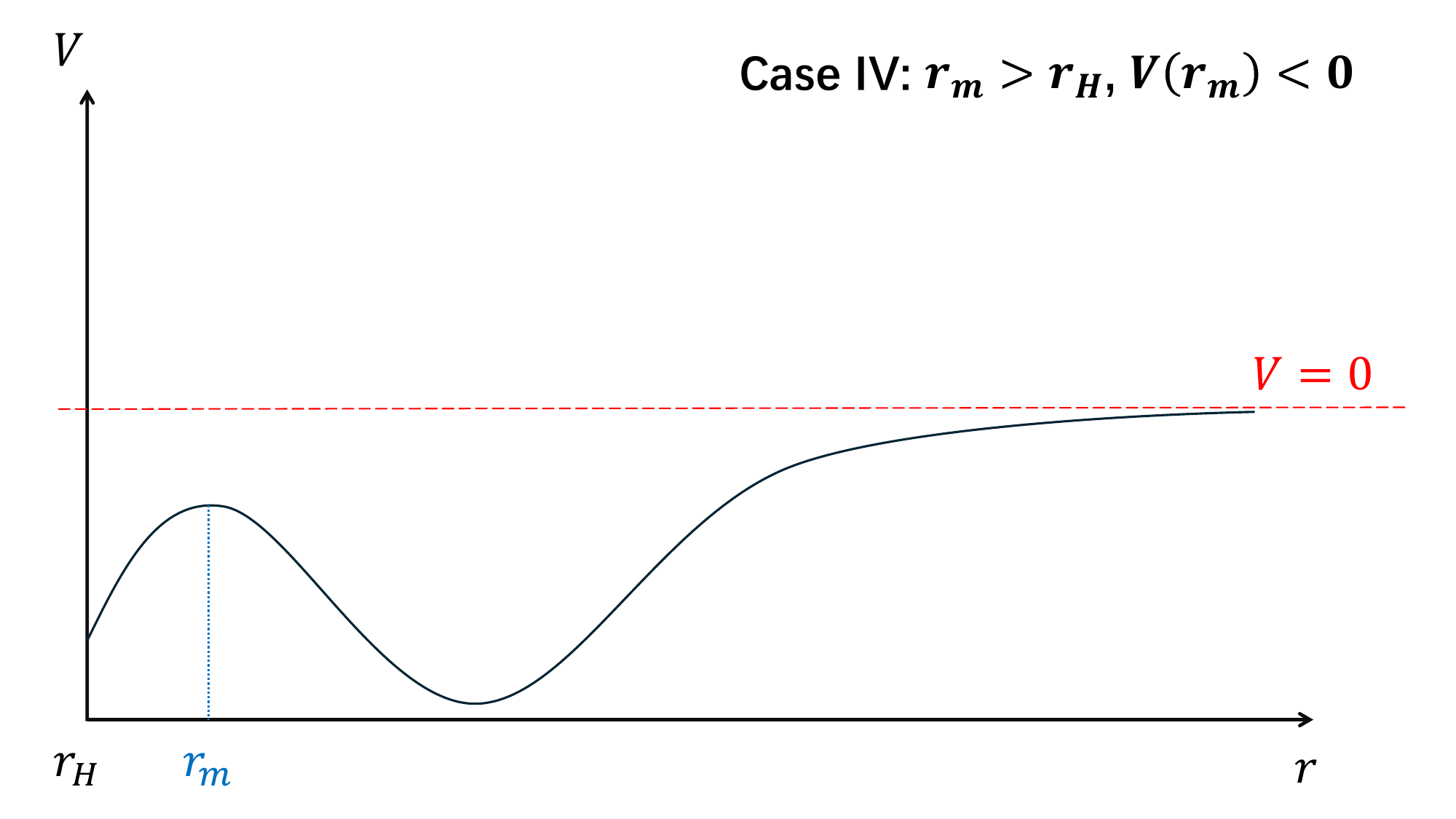}
    \caption{Four cases of effective potential (when the roots of $V'(r)$ are both real) that the generated plasmoids may encounter.}
    \label{fig:cases}
\end{figure*}

As introduced in Sect.~\ref{sec:intro}, reconnections occur frequently in regions where the bulk plasma is sparse, typically in the vicinity of the central black hole. Moreover, the plasmoids are nearly neutral and demagnetized. It is therefore reasonable to presume that the plasmoids move along geodesics once ejected. For equatorial geodesics, for which the Carter constant vanishes, there are two conserved quantities: the energy-at-infinity density $\mathcal{E}\equiv u^{\rm p,0}_t$ and the angular-momentum-at-infinity density $\mathcal{L}\equiv u^{\rm p,0}_{\phi}$ (here "$\pm$" is omitted for brevity). They are determined by the bulk plasma streamline and magnetization, the radial position of the reconnection point, and the orientation angle of the reconnection. The radial equation of motion for a plasmoid can be written as:
\begin{equation}
    \frac{1}{2}\left(\frac{dr}{d\tau}\right)^2 = K-V(r)
    \label{eq:eom}
\end{equation}
where
\begin{equation}
    K := \frac{\mathcal{E}^2-1}{2}
    \label{eq:K}
\end{equation}
is the effective kinetic energy, and
\begin{equation}
    V(r) := -\frac{1}{r}+\frac{\mathcal{L}^2-a^2\left(\mathcal{E}^2-1\right)}{2r^2}-\frac{\left(\mathcal{L}-a\mathcal{E}\right)^2}{r^3}
    \label{eq:V}
\end{equation}
is the effective potential for radial motion. Once a plasmoid is ejected, its effective kinetic energy $K$ keeps constant. Whether it can escape to infinity depends on the value of $K$ and the behavior of its effective potential $V(r)$. 

\begin{table}[htbp]
    \centering
    \resizebox{0.48\textwidth}{!}{
    \begin{tabular}{c|c|c|c|c}
        \hline
        $V(r)$ & $u^r_{\rm p,0}$ & $r_0$ & $K$ & Destiny  \\
        \hline
        \hline
        \multirow{3}{*}{Case 0} & \multirow{2}{*}{$> 0$} & \multirow{2}{*}{~} & $> 0$ & escape \\
        \cline{4-5}
         & & & $< 0$ & BH \\
        \cline{2-5}
         & $< 0$ & ~ & ~ & BH \\
        \hline
        \hline
        \multirow{8}{*}{Case I} & \multirow{4}{*}{$> 0$} & \multirow{2}{*}{$> r_m$} & $> 0$ & escape \\
        \cline{4-5}
         & & & $< 0$ & bounded \\
        \cline{3-5}
         & & \multirow{2}{*}{$< r_m$} & $> V(r_m)$ & escape \\
        \cline{4-5}
         & & & $< V(r_m)$ & BH \\
        \cline{2-5}
         & \multirow{4}{*}{$< 0$} & \multirow{3}{*}{$> r_m$} & $> V(r_m)$ & BH \\
        \cline{4-5}
         & & & $0\sim V(r_m)$ & escape \\
        \cline{4-5}
         & & & $< 0$ & bounded \\
        \cline{3-5}
         & & $< r_m$ & ~ & BH \\
        \hline
        \hline
        \multirow{5}{*}{Case II} & \multirow{2}{*}{$> 0$} & \multirow{2}{*}{~} & $> 0$ & escape \\
        \cline{4-5}
         & & & $< 0$ & bounded \\
        \cline{2-5}
         & \multirow{3}{*}{$< 0$} & \multirow{3}{*}{~} & $> V(r_H)$ & BH \\
        \cline{4-5}
         & & & $0\sim V(r_H)$ & escape \\
        \cline{4-5}
         & & & $< 0$ & bounded \\
        \hline
        \hline
        \multirow{5}{*}{Case III} & \multirow{3}{*}{$> 0$} & \multirow{3}{*}{~} & $> 0$ & escape \\
        \cline{4-5}
         & & & $V(r_H)\sim 0$ & BH \\
        \cline{4-5}
         & & & $< V(r_H)$ & bounded \\
        \cline{2-5}
         & \multirow{2}{*}{$< 0$} & \multirow{2}{*}{~} & $> V(r_H)$ & BH \\
        \cline{4-5}
         & & & $< V(r_H)$ & bounded \\
        \hline
        \hline
        \multirow{8}{*}{Case IV} & \multirow{5}{*}{$> 0$} & \multirow{3}{*}{$> r_m$} & $> 0$ & escape \\
        \cline{4-5}
         & & & $V(r_m)\sim 0$ & BH \\
        \cline{4-5}
         & & & $< V(r_m)$ & bounded \\
        \cline{3-5}
         & & \multirow{2}{*}{$<r_m$} & $< 0$ & BH \\
        \cline{4-5}
         & & & $> 0$ & escape \\
        \cline{2-5}
         & \multirow{3}{*}{$< 0$} & \multirow{2}{*}{$> r_m$} & $> V(r_m)$ & BH \\
        \cline{4-5}
         & & & $< V(r_m)$ & bounded \\
        \cline{3-5}
         & & $< r_m$ & ~ & BH \\
        \hline
    \end{tabular}
    }
    \caption{Destinies of ejected plasmoids for different situations.}
    \label{tab:destiny}
\end{table}

From Eq.~\eqref{eq:V}, it is apparent that $V(r)$ approaches 0 as $r\rightarrow \infty$ and it has generally two types of shapes. In the first type, $V(r)$ increases monotonically with $r$ (referred to as \textbf{Case 0}), for which the equation $V'(r)=0$ has no real root or has a real double root. In the second type, $V(r)$ has a maximal value at $r_m$ (satisfying $V'(r_m)=0$ and $V''(r_m)<0$) and a minimal value at $\Tilde{r}_m$ (satisfying $V'(\Tilde{r}_m)=0$ and $V''(\Tilde{r}_m)>0$). There are totally four distinct cases (called \textbf{Case I-IV}) we should consider for the second type, depending on the position of $r_m$ and the sign of $V(r_m)$. These cases are illustrated in Fig.~\ref{fig:cases}.

We trace the following steps to determine the destiny of an ejected plasmoid:
\begin{itemize}
    \item Step 1: Determine the shape of $V(r)$.
    \item Step 2: Determine the sign of $u_{\rm p,0}^r$, namely determine whether the plasmoid is shot radially inward or outward from the reconnection point.
    \item Step 3: Compare the radial position of the reconnection point with characteristic radii, including $r_m$ and the radius of event horizon $r_H$.
    \item Step 4: Compare the value of $K$ with the relevant extrema of $V(r)$.
\end{itemize}
Upon completing these four steps, the destiny of an ejected plasmoid can be unambiguously determined. The outcomes for all of the cases are summarized in Tab.~\ref{tab:destiny}.

Besides, there is one special case which is not listed in Tab.~\ref{tab:destiny}. It is probable that $u^r_{\rm p,0}=0$, meaning that the plasmoid is thrown into a circular orbit. In this case, if $r_0=r_m$, the orbit is unstable and the plasmoid may eventually escape; if $r_0=\Tilde{r}_m$, the orbit is stable and the plasmoid remains bounded.

\section{Results}
\label{sec:results}

\subsection{Destinies of $p^+$ Shown in $\xi_{B}-r_0$ Parameter Planes}
\label{sec:para-planes}

\begin{figure*}
    \centering
    \includegraphics[width=0.32\linewidth]{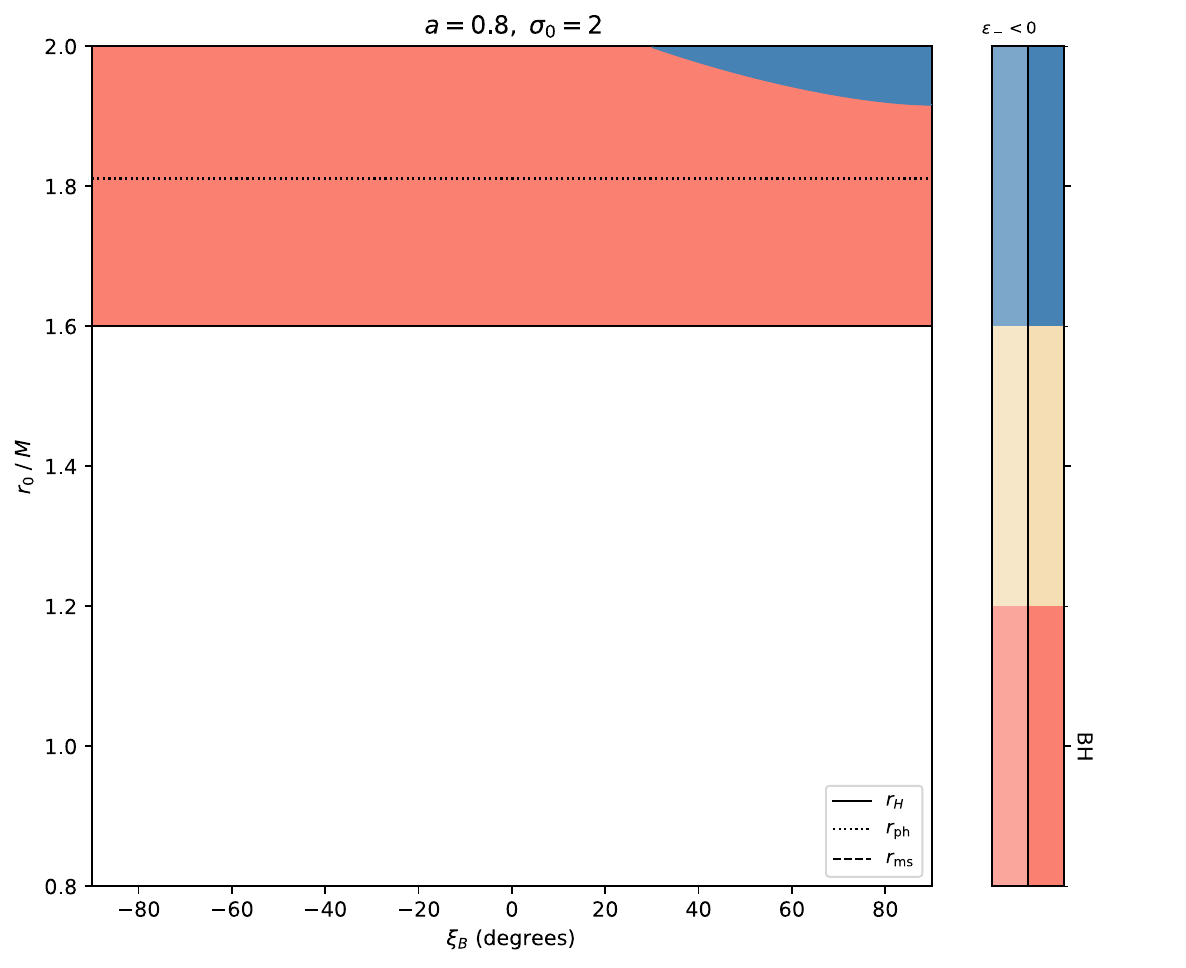}
    \includegraphics[width=0.32\linewidth]{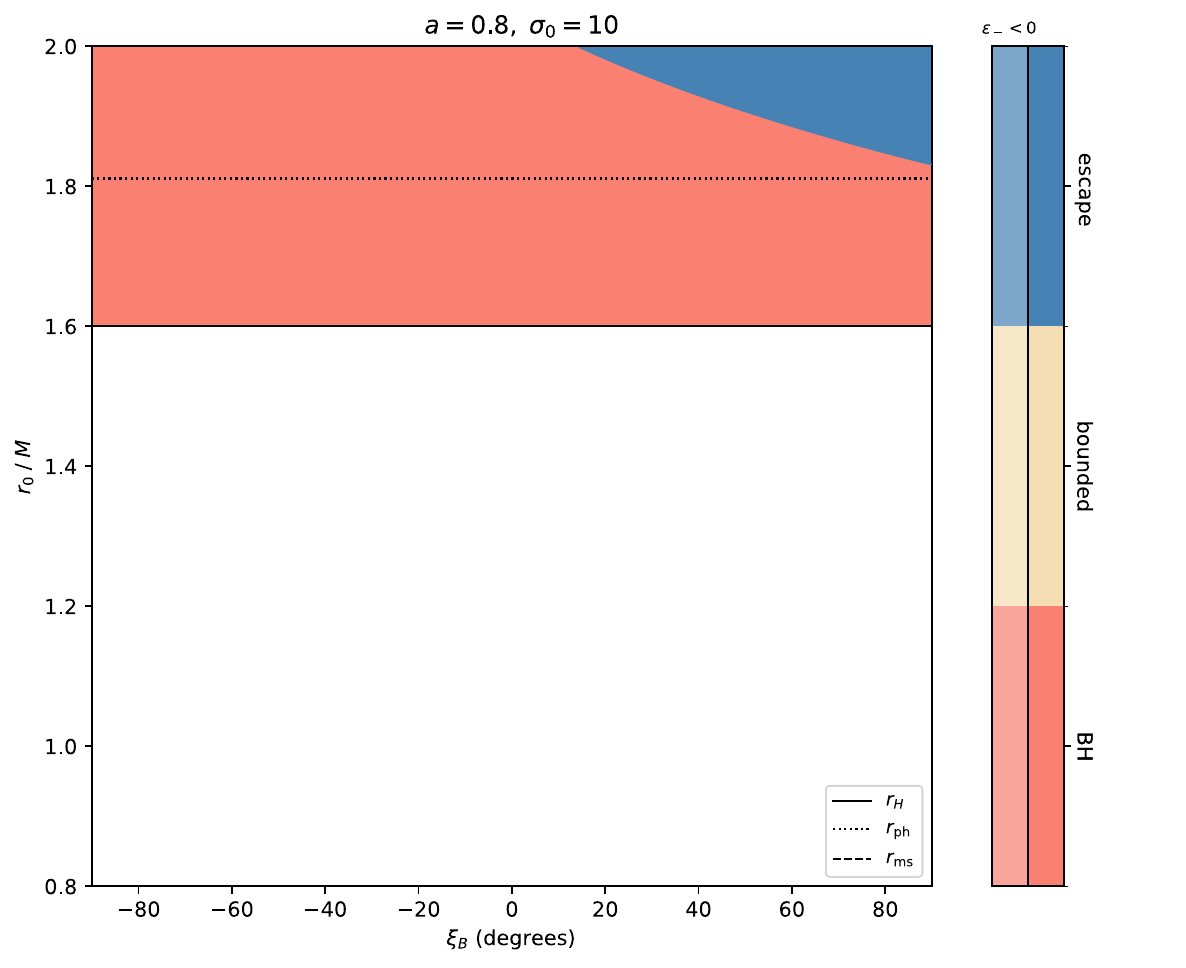}
    \includegraphics[width=0.32\linewidth]{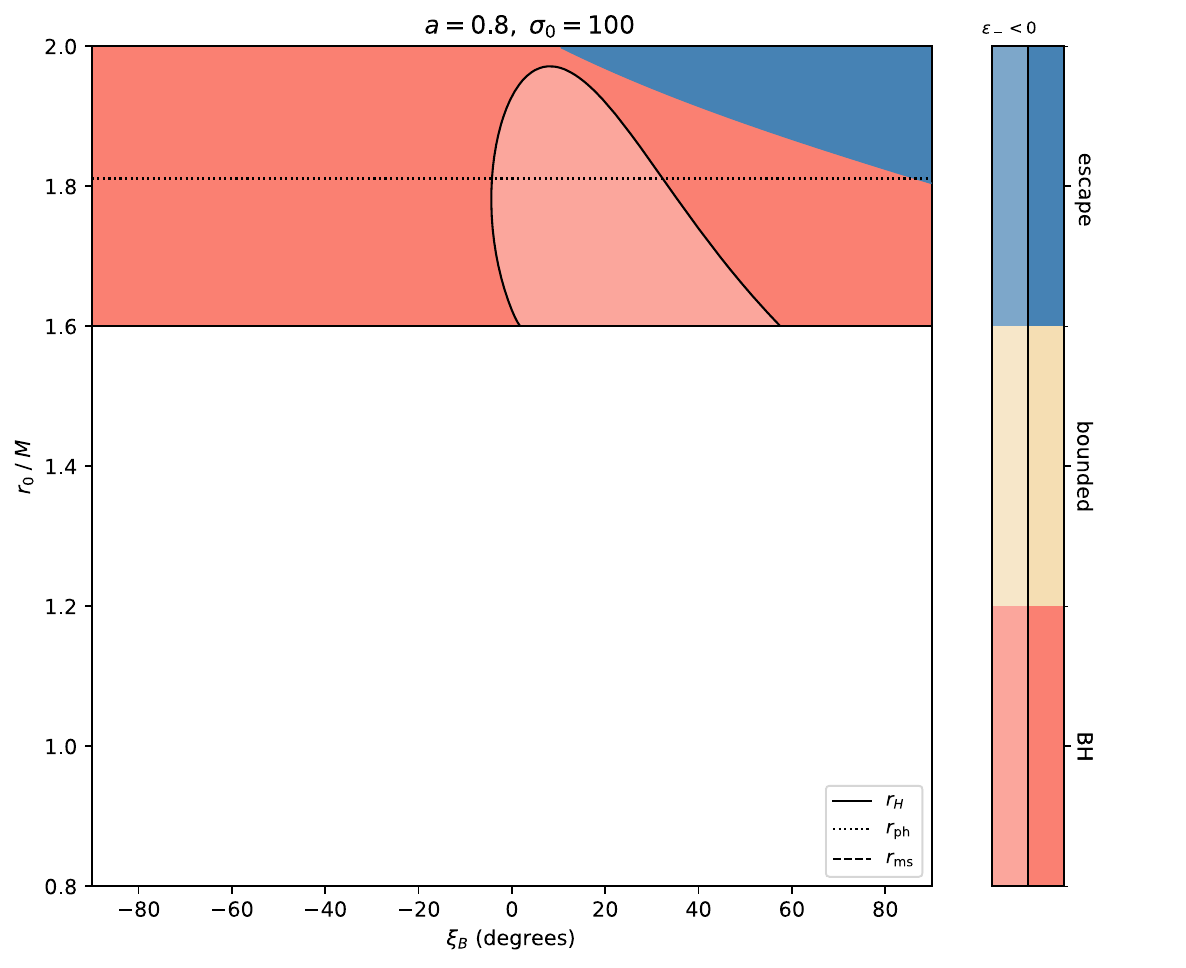}
    \\
    ~~~~~~~~~~~~~~~~~~~~~~
    \\
    \includegraphics[width=0.32\linewidth]{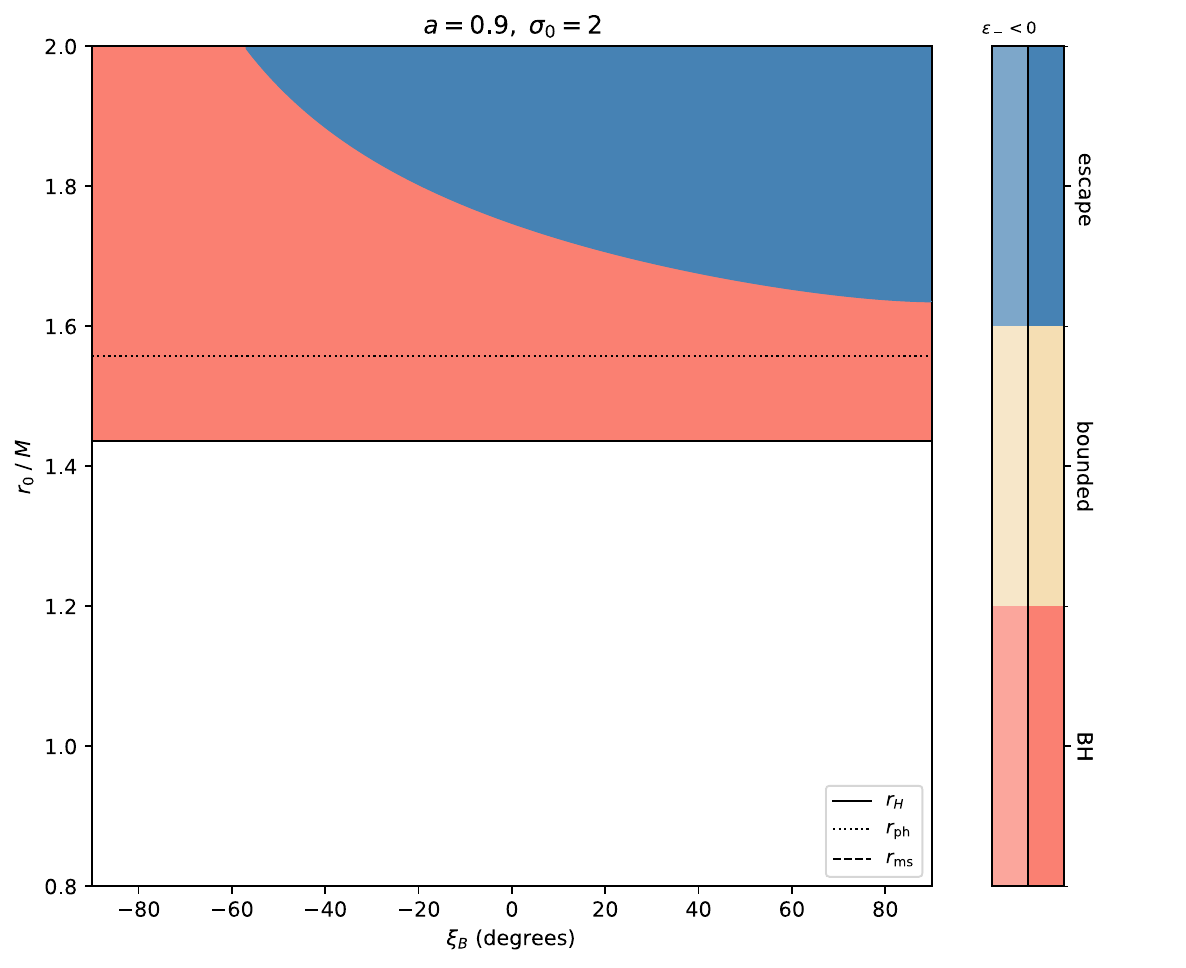}
    \includegraphics[width=0.32\linewidth]{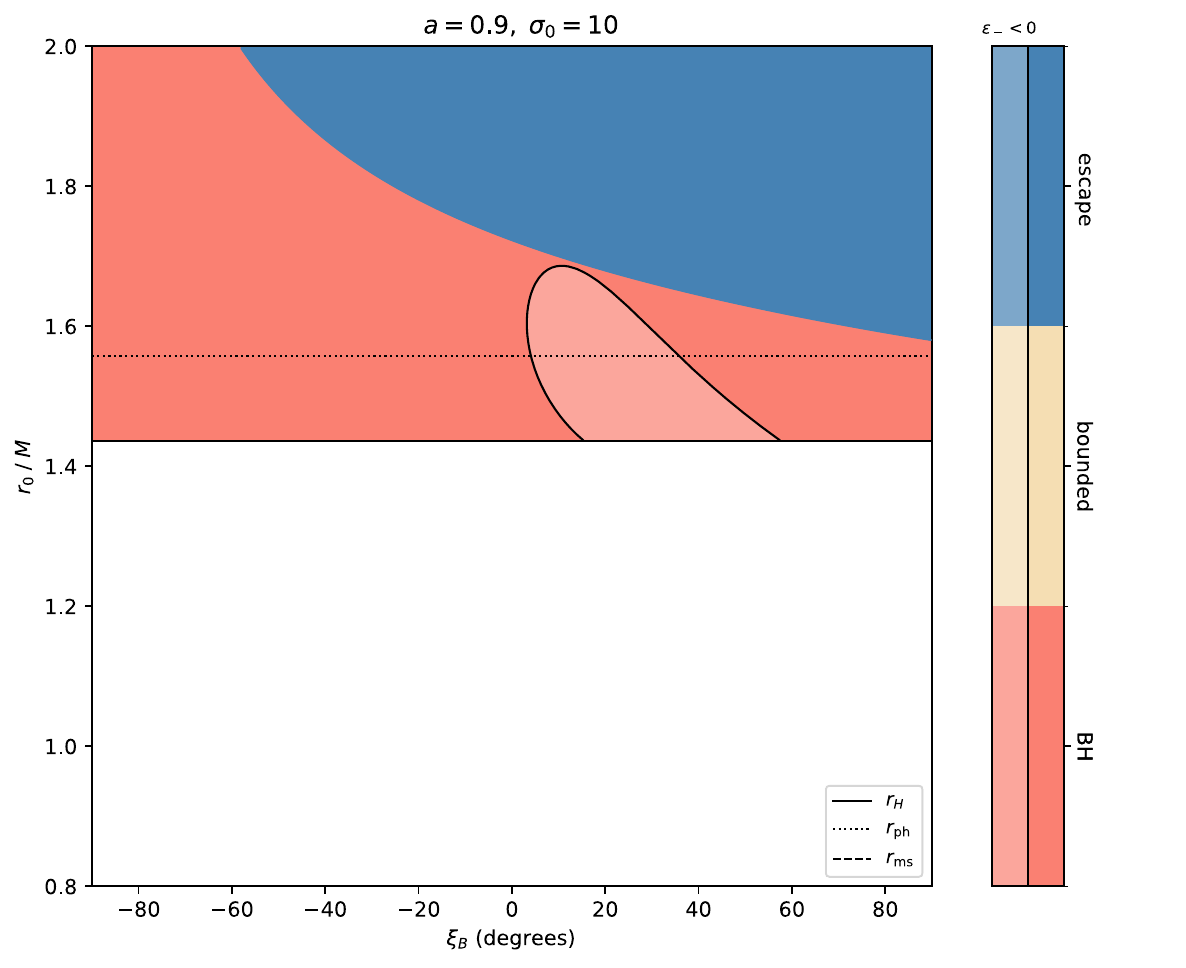}
    \includegraphics[width=0.32\linewidth]{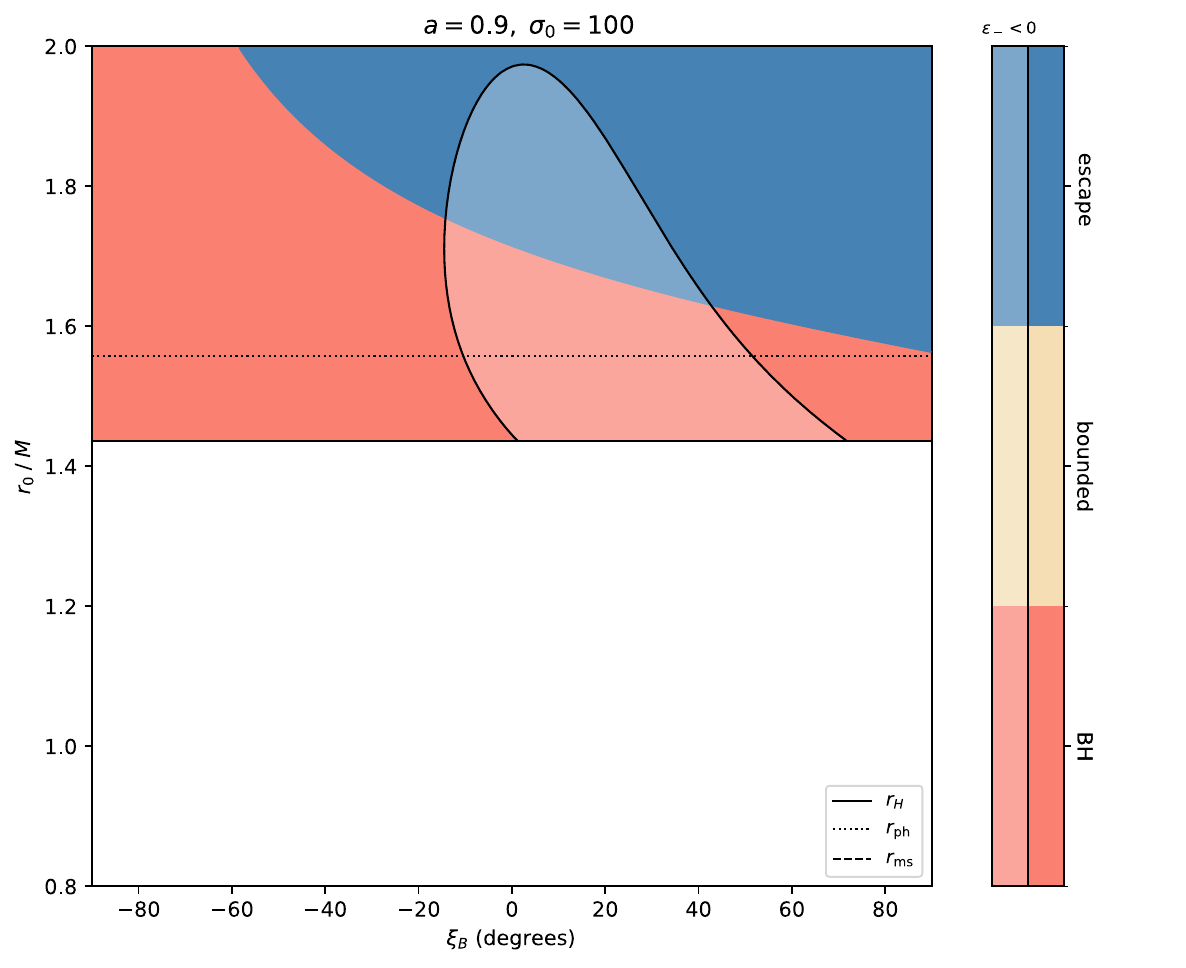}
    \\
    ~~~~~~~~~~~~~~~~~~~~~~
    \\
    \includegraphics[width=0.32\linewidth]{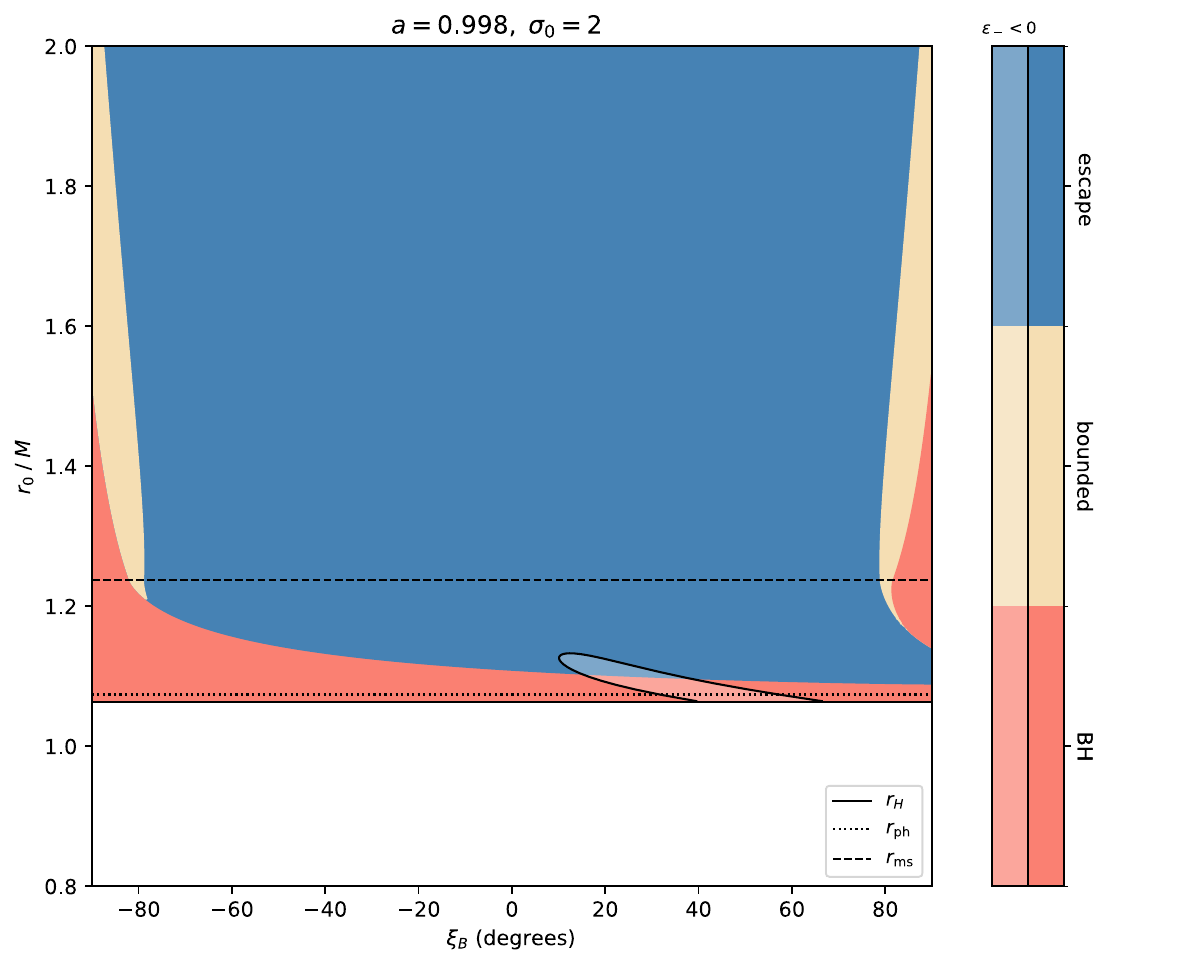}
    \includegraphics[width=0.32\linewidth]{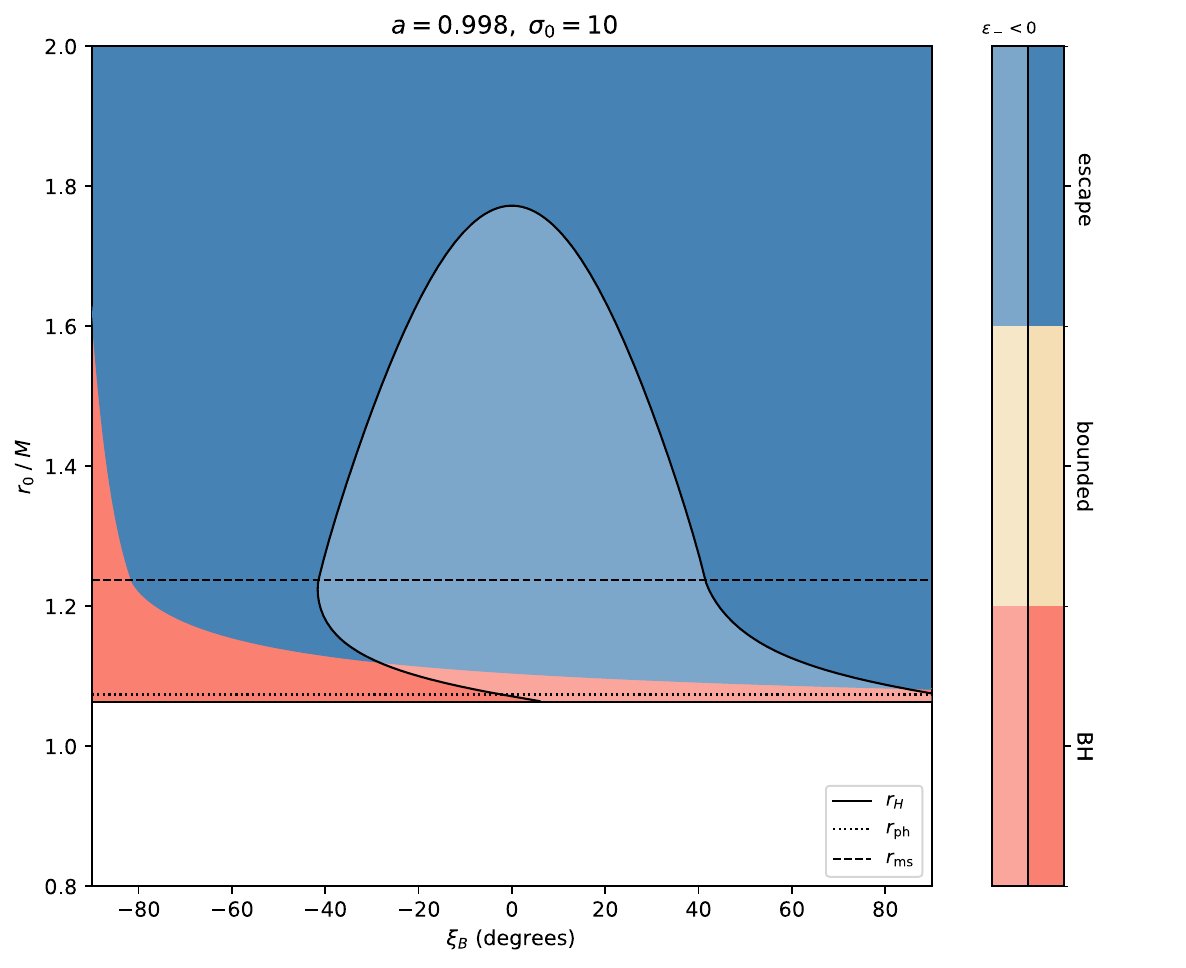}
    \includegraphics[width=0.32\linewidth]{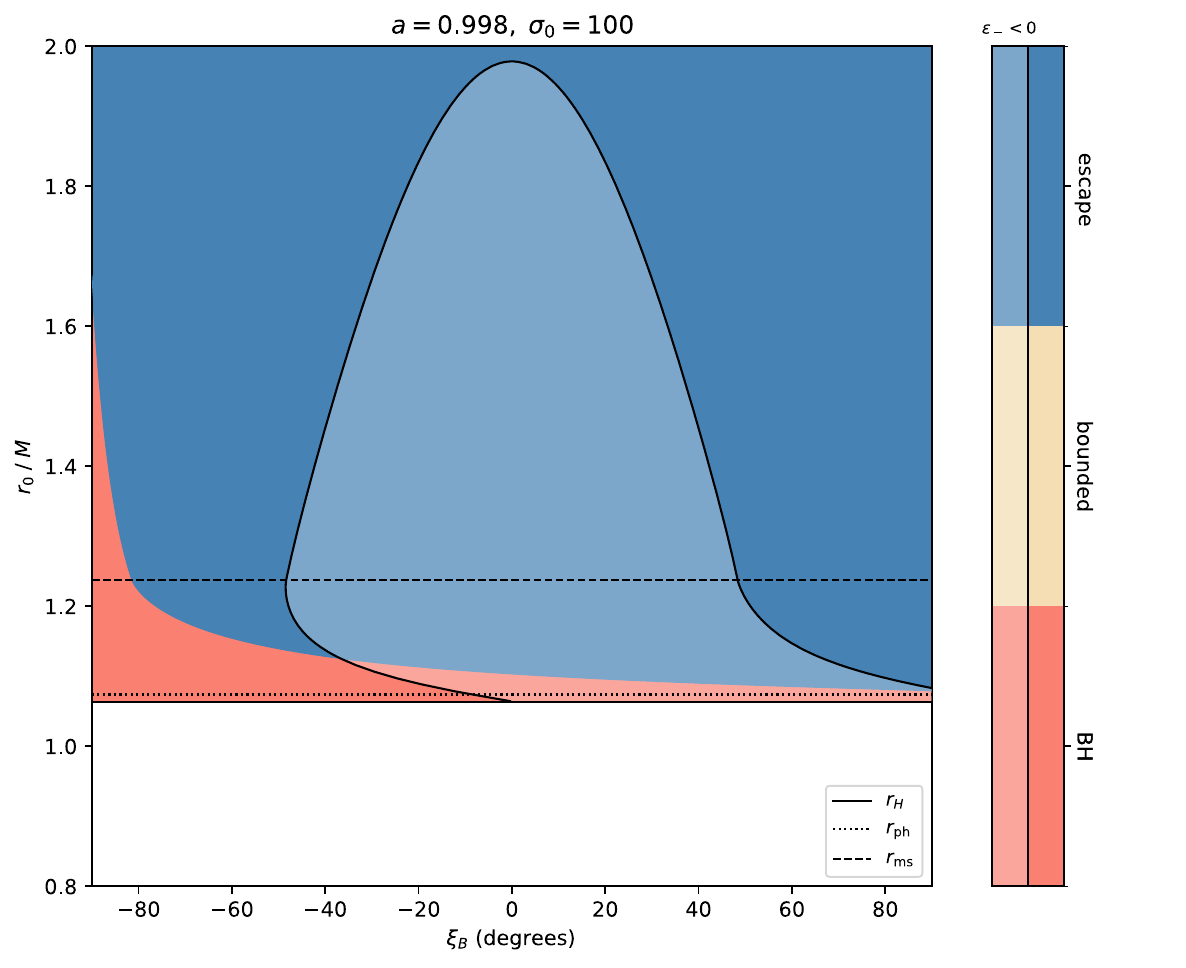}
    \caption{The $\xi_B-r_0$ parameter planes on which the escaping p$^+$, bounded p$^+$ and those that plunge into the central black hole are colored in blue, yellow and red, respectively. The boundaries of energy extraction regions are represented by solid curves. The dotted and dashed straight lines represent the photon sphere and ISCO, respectively. The left, central and right panels exhibit the results of $\sigma_0=$ 2, 10 and 100, while the top, middle and bottom panels exhibit the results of $a=$ 0.8, 0.9, 0.998, respectively.}
    \label{fig:xi-r0}
\end{figure*}

We adopt a combined streamline for the bulk plasma, which presumes a circularly flowing and plunging bulk plasma outside and inside the innermost stable circular orbit (ISCO), respectively \cite{work1,Mummery:2022ana}. This type of streamline is considered more realistic for astrophysical accretion systems, and the plunging region is thought to account for the presence of soft X-ray excess \cite{Mummery:2023tgh,Mummery:2024tyo}. In Fig.~\ref{fig:xi-r0}, we present the destinies of p$^+$ after formation on the $\xi_B-r_0$ parameter planes, where $\xi_B$ ranges from $-\pi/2$ to $\pi/2$ and $r_0$ ranges from the radius of event horizon to that of the ergosphere. The boundaries of energy extraction regions, the parameter domains in which p$^-$ with negative energy-at-infinity density are generated via magnetic reconnection, are delineated by solid curves. We may as well refer these regions to be energy extraction region for escaping p$^+$, within which p$^+$ escapes to the infinity with its corresponding p$^-$ carrying negative energy-at-infinity density.

We can see from Fig.~\ref{fig:xi-r0} that the generated p$^+$ are rarely bounded. Almost all p$^+$ either escape to infinity or are absorbed by the central black hole. When the central black hole rotates relatively slowly, like $a=0.8$ as the top panels of Fig.~\ref{fig:xi-r0} show, most p$^+$ are absorbed, with the exception of those ejected radially outward from reconnection points located sufficiently far from the event horizon. The p$^-$ with negative energy-at-infinity density are formed only when the magnetization is extremely large, as shown in the top right panel, corresponding to p$^+$ carrying the energy extracted from the central black hole. Nevertheless, energy extraction via magnetic reconnection would never succeed because the corresponding p$^+$ fail to escape. 

For relatively fast black hole rotation, like $a=0.9$ as the middle panels of Fig.~\ref{fig:xi-r0} show, nearly half of p$^+$ could escape after ejection. However, successful energy extraction via magnetic reconnection still demands extremely high magnetization of bulk plasma ($\sigma_0\sim 100$). When the black hole spin approaches extreme, like $a=0.998$ as the bottom panels of Fig.~\ref{fig:xi-r0} show, most p$^+$ escape to infinity after formation, except those generated very close to the event horizon. This outcome is expected, as the strong centrifugal effect induced by rapid black hole rotation tends to propel the plasmoids outward. In this situation, the destinies of ejected p$^+$ need not to be a major concern. A moderately high magnetization of the bulk plasma ($\sigma_0\sim 10$) suffices to produce a substantial number of escaping p$^+$ carrying energy extracted from the central black hole. The area of energy extraction region and that of energy extraction region for escaping p$^+$ are found to be nearly identical. 

\subsection{Covering Factor for Escaping $p^+$}
\label{sec:covering}

In Ref.~\cite{work1,work2}, the covering factor for energy extraction $\chi$ is defined as the ratio of the area of the energy extraction region to the total area of the $\xi_B-r_0$ parameter plane. Based on the discussions of the destinies of generated p$^+$, we further define $\Tilde{\chi}$, the covering factor for escaping p$^+$, as the ratio of the area of the energy extraction region for escaping p$^+$ to the total area of the $\xi_B-r_0$ parameter plane. The covering factor for escaping p$^+$, as an intrinsic property, reflects the capability of an accretion system, whose magnetization is assumed to distribute uniformly, to extract energy via magnetic reconnection from its central black hole, with only escaping p$^+$ being counted. A higher $\Tilde{\chi}$ indicates a higher probability of successful energy extraction for an accretion system with uniform magnetization. Namely, a greater fraction of plasmoids formed from magnetic reconnections contain energy extracted from the central black hole and manage to reach infinity for a uniformly magnetized accretion system with a higher $\Tilde{\chi}$.

\begin{figure*}
    \centering
    \includegraphics[width=\linewidth]{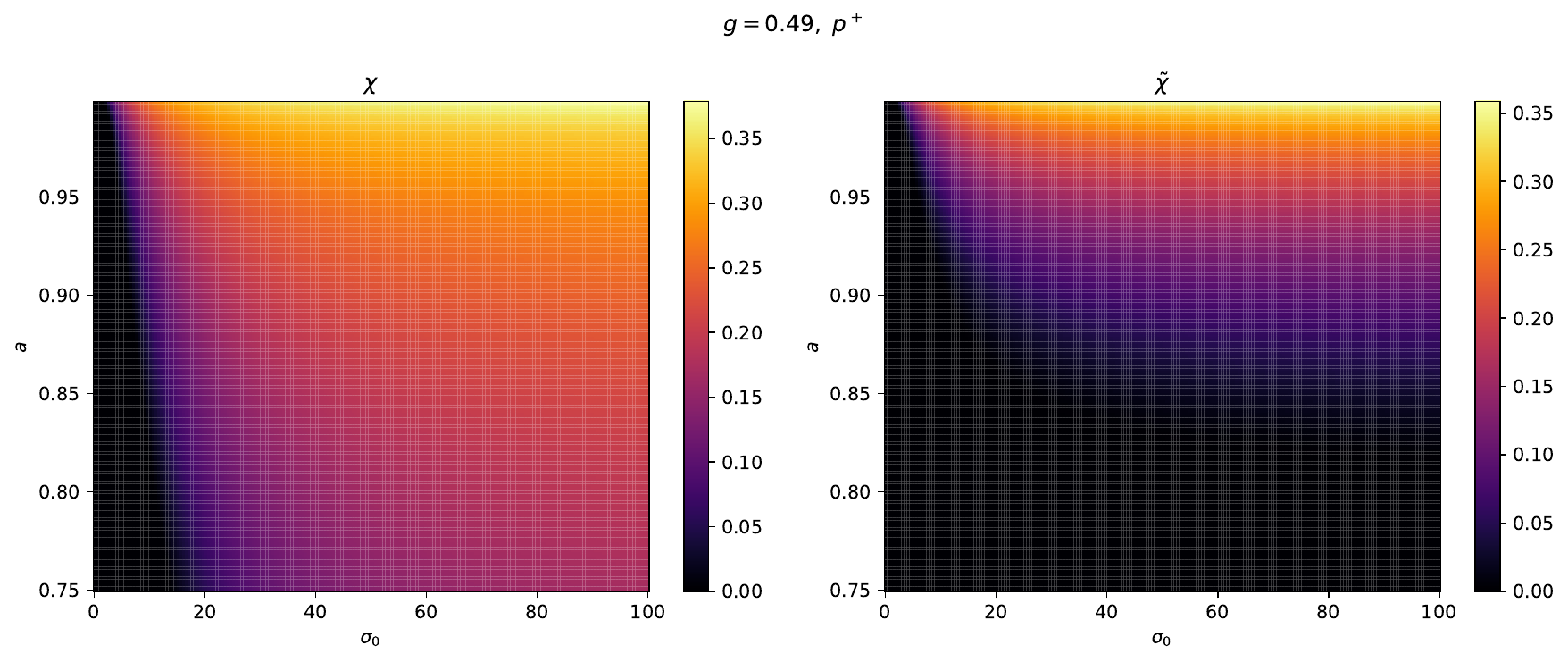}
    \caption{The distributions of $\chi$ (left panel) and $\Tilde{\chi}$ (right panel) on $\sigma_0-a$ parameter planes.}
    \label{fig:a-sig}
\end{figure*}

\begin{figure*}
    \centering
    \includegraphics[width=0.45\linewidth]{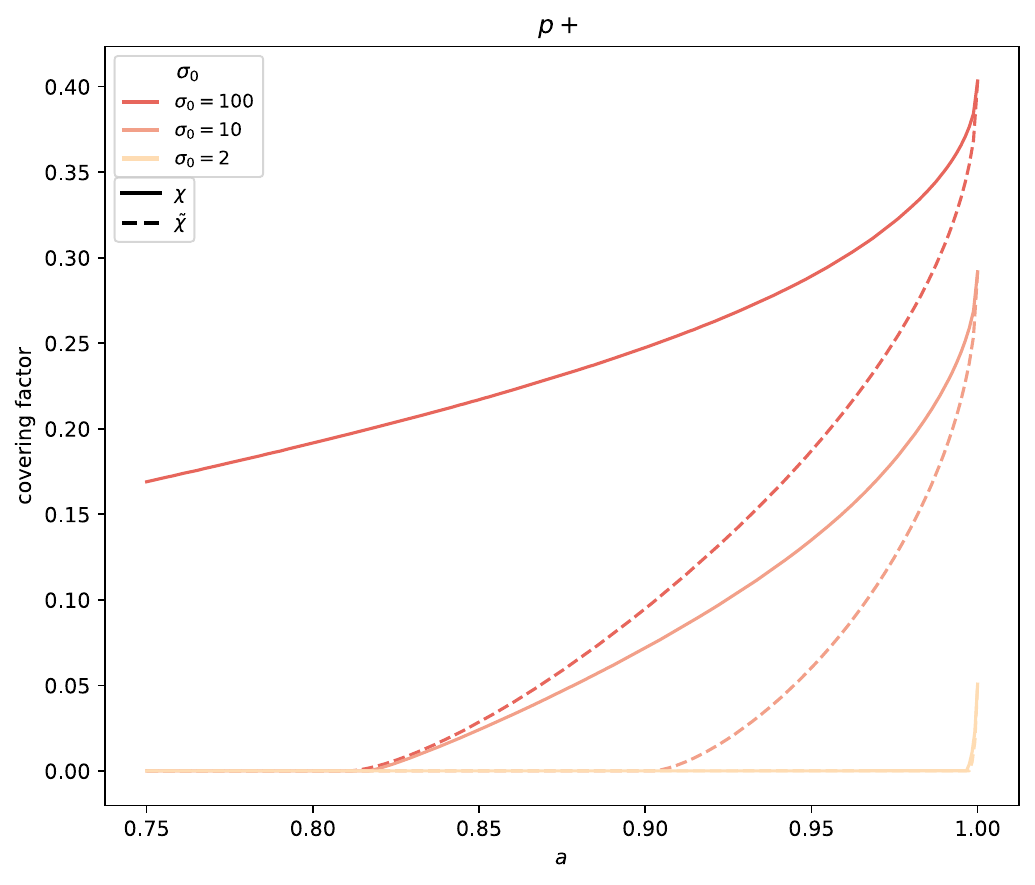}
    \includegraphics[width=0.45\linewidth]{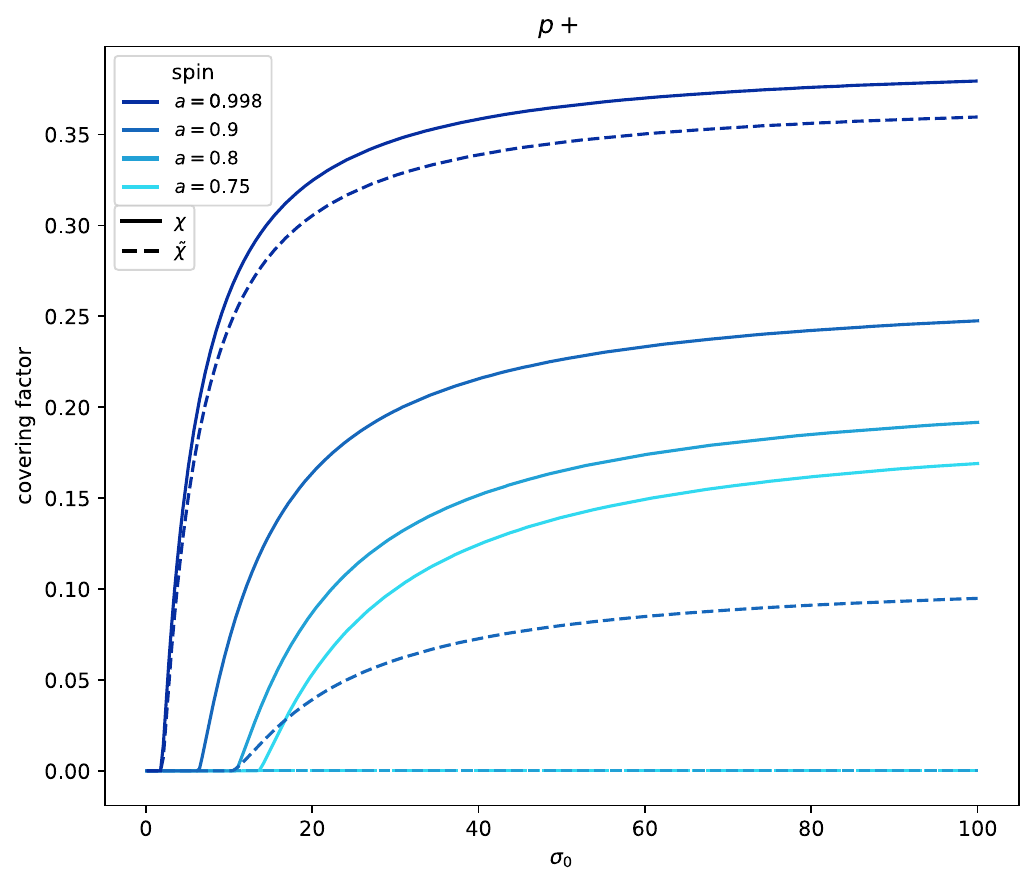}
    \caption{Left panel: covering factors $\chi$ and $\Tilde{\chi}$ as functions of black hole spin $a$. Right panel: of magnetization $\sigma_0$.}
    \label{fig:chi-a-sig}
\end{figure*}

The distributions of $\chi$ and $\Tilde{\chi}$ on $\sigma_0-a$ parameter planes are plotted on the left and right panels of Fig.~\ref{fig:a-sig}, respectively. Meanwhile, we plot the variations of $\chi$ and $\Tilde{\chi}$ with $a$ for multiple values of $\sigma_0$ on the left panel of Fig.~\ref{fig:chi-a-sig} and the variations of $\chi$ and $\Tilde{\chi}$ with $\sigma_0$ for multiple values of $a$ on the right panel of Fig.~\ref{fig:chi-a-sig}. It is apparent that $\chi$ is generally greater than $\Tilde{\chi}$. For relatively low black hole spin ($a\lesssim 0.9$), the discrepancy between $\chi$ and $\Tilde{\chi}$ remains significant even at high magnetization ($\sigma_0\gtrsim 100$). In this case, a considerable fraction of ejected p$^+$ are ultimately captured by the central black hole, as the centrifugal effect is insufficient to propel them outward. 

As the black hole spin approaches extreme ($a\rightarrow 1$), however, the difference between $\chi$ and $\Tilde{\chi}$ becomes negligible, as Fig.~\ref{fig:a-sig} and Fig.~\ref{fig:chi-a-sig} illustrate. In this situation, the centrifugal effect is so strong that even plasmoids ejected with moderate speeds ($v_{\rm out}\sim 0.3$) can readily escape. Furthermore, in the case of extreme black hole spin, the influence of magnetization on $\Tilde{\chi}$ is almost negligible once it is sufficiently high ($\sigma_0\gtrsim 20$). In short, the black hole spin plays a more crucial role than the bulk plasma magnetization for the escape of p$^+$ produced by magnetic reconnections, and therefore is more important for the overall success of energy extraction from the central black hole.

\begin{figure*}
    \centering
    \includegraphics[width=0.45\linewidth]{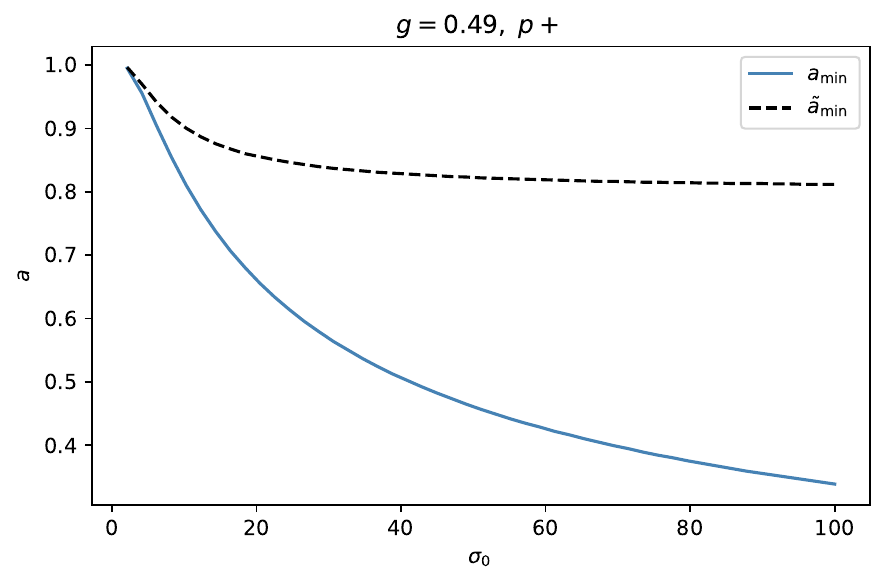}
    \includegraphics[width=0.45\linewidth]{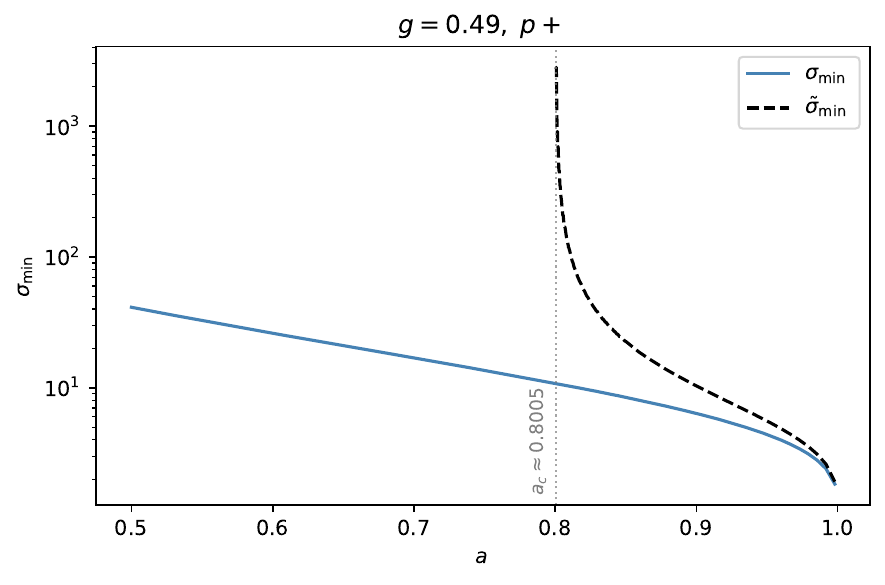}
    \caption{Left panel: $a_{\rm min}$ (solid line) and $\Tilde{a}_{\rm min}$ (dashed line) as functions of $\sigma_0$. Right panel: $\sigma_{\rm min}$ (solid line) and $\Tilde{\sigma}_{\rm min}$ (dashed line) as functions of $a$. The grey dotted line represents $a=0.8005$.}
    \label{fig:chi-a-chi-sig}
\end{figure*}

One may define the minimum allowed black hole spin for energy extraction at a given magnetization, or conversely the minimum allowed magnetization of bulk plasma for energy extraction for a given black hole spin, by imposing $\chi=0$ or $\Tilde{\chi}=0$. The variations of $a_{\rm min}$ (for $\chi=0$) and $\Tilde{a}_{\rm min}$ (for $\Tilde{\chi}=0$) with $\sigma_0$ are shown on the left panel of Fig.~\ref{fig:chi-a-chi-sig}. It is evident that relatively low black hole spins are permitted for successful energy extraction if the destinies of the plasmoids are ignored. However, the behavior of $\Tilde{a}_{\rm min}$ reveals that the energy extraction would completely succeed, for which the plasmoid carrying extracted energy actually escapes to infinity, only for sufficiently high black hole spins. This point is more clearly illustrated on the right panel of Fig.~\ref{fig:chi-a-chi-sig}, which displays the variations of $\sigma_{\rm 0,min}$ (for $\chi=0$) and $\Tilde{\sigma}_{\rm 0,min}$ (for $\Tilde{\chi}=0$) as functions of $a$, where the $\sigma_0$-axis is in logarithmic scale for easy inspection of the high magnetization regime. It implies that $a\approx 0.8005$ is a fundamental lower bound on the black hole spin for successful energy extraction if the escape of p$^+$ that carries the extracted energy is required, regardless of how large the bulk plasma magnetization may be. In fact, this limit can be obtained directly by taking $v_{\rm out}\rightarrow 1$ in Eq.~\eqref{eq:u_out}. In other words, the successful energy extraction is impossible when the central black hole spin is lower than 0.8005, for that the plasmoid carrying the extracted energy could never reach infinity. As is well known, magnetic reconnection would successively occur in the ergo region of a rotating black hole and the generations of p$^-$ with negative energy-at-infinity density attenuate the spin of the central black hole. However, based on our findings, this spin-down process would cease once the black hole spin drops below 0.8005.

\section{Summary and Discussions}
\label{sec:sum}

For complete energy extraction from a rotating black hole via magnetic reconnection within the ergosphere, two conditions have to be satisfied simultaneously. First, one of the ejected plasmoids, denoted as p$^-$, must carry negative energy-at-infinity density. Second, the other plasmoid, denoted as p$^+$, must overcome the gravitational pull of the black hole and escape to infinity. The first condition has already been extensively examined under various scenarios in previous works. The second condition, however, is the central focus of the present study.

We presume that the plasmoids follow geodesic motion after ejection, since they are nearly neutral, demagnetized and generated in the region where the bulk plasma is sparse. By analyzing the effective kinetic energy and the behavior of effective potential of radial motion, we determine the destinies of plasmoids (escape, bounded or absorbed by the central black hole) across all relevant situations. We exhibit the destinies of p$^+$ on $\xi_B-r_0$ parameter planes for multiple values of magnetization and black hole spin. By plugging the energy extraction regions (obtained in previous works) into the $\xi_B-r_0$ parameter planes, we acquire the energy extraction region for escaping p$^+$, wherein complete energy extraction from the central black hole via magnetic reconnection is feasible.

The covering factor $\chi$ was defined to quantify the capability of an accretion system, whose magnetization is assumed to distribute uniformly, to extract energy from its central black hole in previous works \cite{work1,work2}. By discussing the destinies of p$^+$, we define $\Tilde{\chi}$, the covering factor for escaping p$^+$, correspondingly. Complete energy extraction could only occur in accretion systems with non-zero $\Tilde{\chi}$. A greater $\Tilde{\chi}$ of an accretion system indicates a higher probability of successful energy extraction. Meanwhile, solving $\Tilde{\chi}=0$ yields the minimum allowed black hole spin $\Tilde{a}_{\rm min}$ or the minimum allowed magnetization of bulk plasma $\Tilde{\sigma}_{\rm 0,min}$ for energy extraction. We find that the difference between $\chi$ and $\Tilde{\chi}$ is inconsiderable when the black hole spin approaches extreme, implying that most p$^+$, carrying energy extracted from the central black hole, eventually escape to infinity. It is a natural consequence since the centrifugal effect is so strong when the central black hole rotates fast. The escape of p$^+$ is generally not a major concern. On the contrary, when the black hole rotates relatively slowly, the requirement of magnetization becomes stringent. More importantly, we identify a critical black hole spin, $a=0.8005$, below which energy extraction in an accretion system from the central black hole could never succeed. As we know, the spin of a rapidly rotating black hole continually decreases when the energy extraction via magnetic reconnection repeatedly occurs. This process would halt when the black hole spin decays to approximately 0.8005, based on our results.

It is also necessary to confirm that p$^-$ with negative energy-at-infinity density plunges into the central black hole before we can assert that energy extraction succeeds completely. We exhibit the destinies of p$^-$ in Fig.~\ref{fig:xi-r0-neg} and it is verified that all p$^-$ are finally absorbed by the central black hole without exception, within the geodesic framework. Moreover, as we know, there are various mechanisms to transport the extracted energy to the infinity, besides the direct escape of p$^+$. For example, p$^+$ radiates the energy and the generated photons escape to the infinity. Additionally, p$^+$ contact the accretion flow and the energy is transported advectively or thermally. What we discussed in this work, the direct escape of p$^+$, is the process that is easiest to consider. Incorporating all these processes is nontrivial, but may hopefully be addressed through numerical simulations in the future.

Up to now, no theoretical model of magnetized accretion flow has been applied in studies of the Comisso-Asenjo process. Though Ref.~\cite{Camilloni:2024tny} applied a magnetized torus model described in Ref.~\cite{Komissarov:2006nz}, that model lacks mass accretion and does not feature an equatorial current sheet. In reality, the magnetizations of bulk plasma should vary with position near the black hole. Consequently, the growth rate of the tearing-mode instability, and thus the probability of magnetic reconnection, will vary throughout the equatorial accretion flow. However, there is still no widely accepted analytical model to handle the distribution of magnetization in accretion flows. This gap might be bridged by combining existing analytical frameworks with results from numerical simulations. The definition of the covering factor should be carefully revised once the Comisso-Asenjo process is examined within a specific magnetized accretion flow model. Its current definition is applicable for a uniformly magnetized accretion flow, which is unnatural in the real black hole accretion system. Only after a feasible analytical model of magnetized accretion flow is proposed to refine the definition of the covering factor can it really describe the possibility of successful energy extraction.

The limit of black hole spin, $a=0.8005$, is obtained by considering the Comisso-Asenjo process only. One cannot conclude that this value represents an equilibrium spin of a rotating black hole in the presence of a surrounding accretion flow where magnetic reconnection occurs repeatedly. Determining such an equilibrium spin under the influence of the accretion system is an important open question, and numerical simulations have already provided some preliminary hints \cite{Jacquemin-Ide:2023aax,Lowell:2023kyu}. Generally speaking, mass accretion and the Blandford-Znajek process \cite{BZ,Menon:2005mg,Davis:2020wea,Camilloni:2022kmx} influence the black hole spin together with the Comisso-Asenjo process. Presumably, the Comisso-Asenjo process dominates only in the non-jetted and highly magnetized accretion, for which the equilibrium spin may be close to the limit obtained in this work.

\begin{acknowledgments}

Ye Shen is supported by NSFC (Grant No.~12605090). Ho-Yun YuChih is supported by NSFC (Grant No.~12275004). Ye Shen would like to thank Dr. Junyi Li (from UMN) for his initial investigation on this topic.

\end{acknowledgments}

\appendix

\section{Destinies of p$^-$}
\label{sec:appendix}

\begin{figure*}
    \centering
    \includegraphics[width=0.32\linewidth]{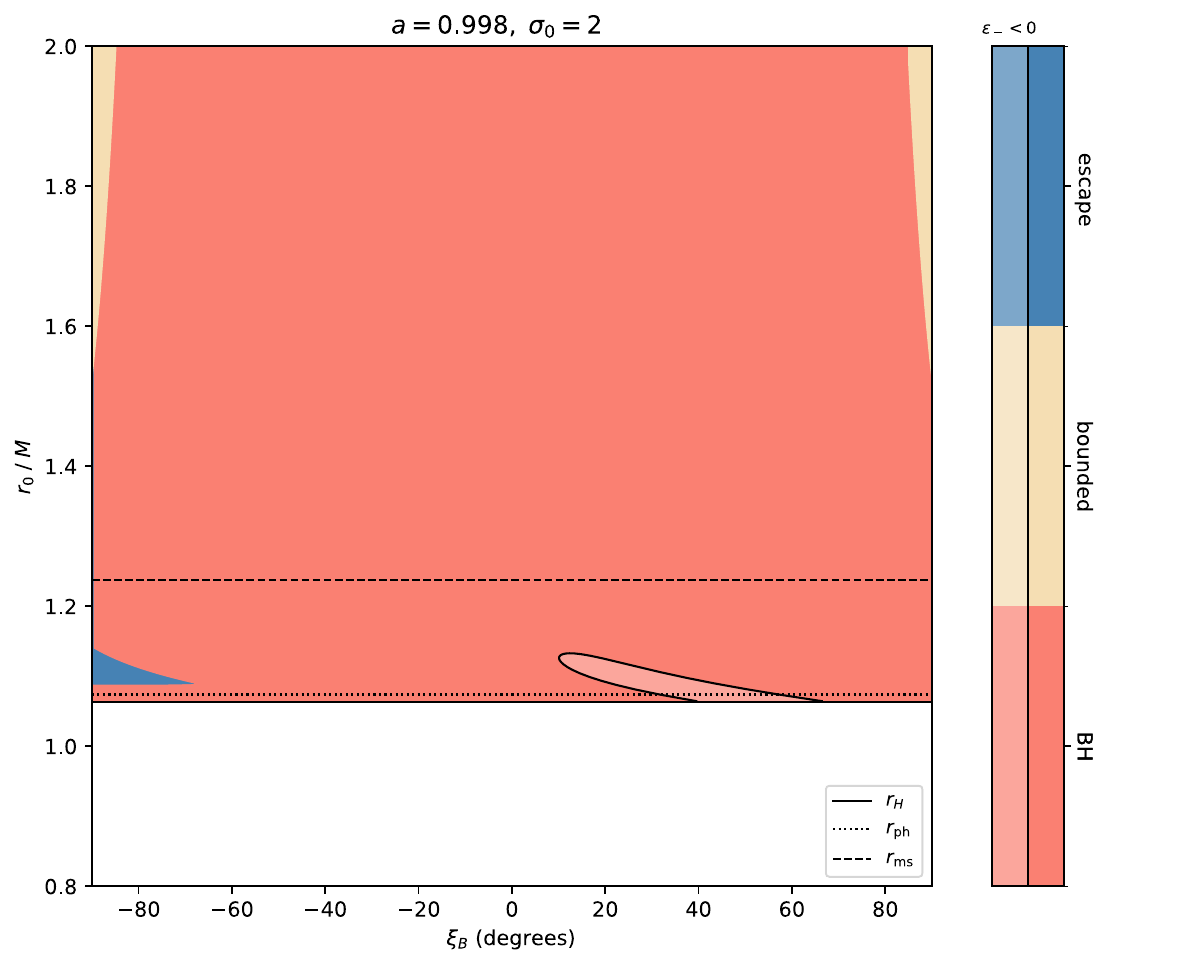}
    \includegraphics[width=0.32\linewidth]{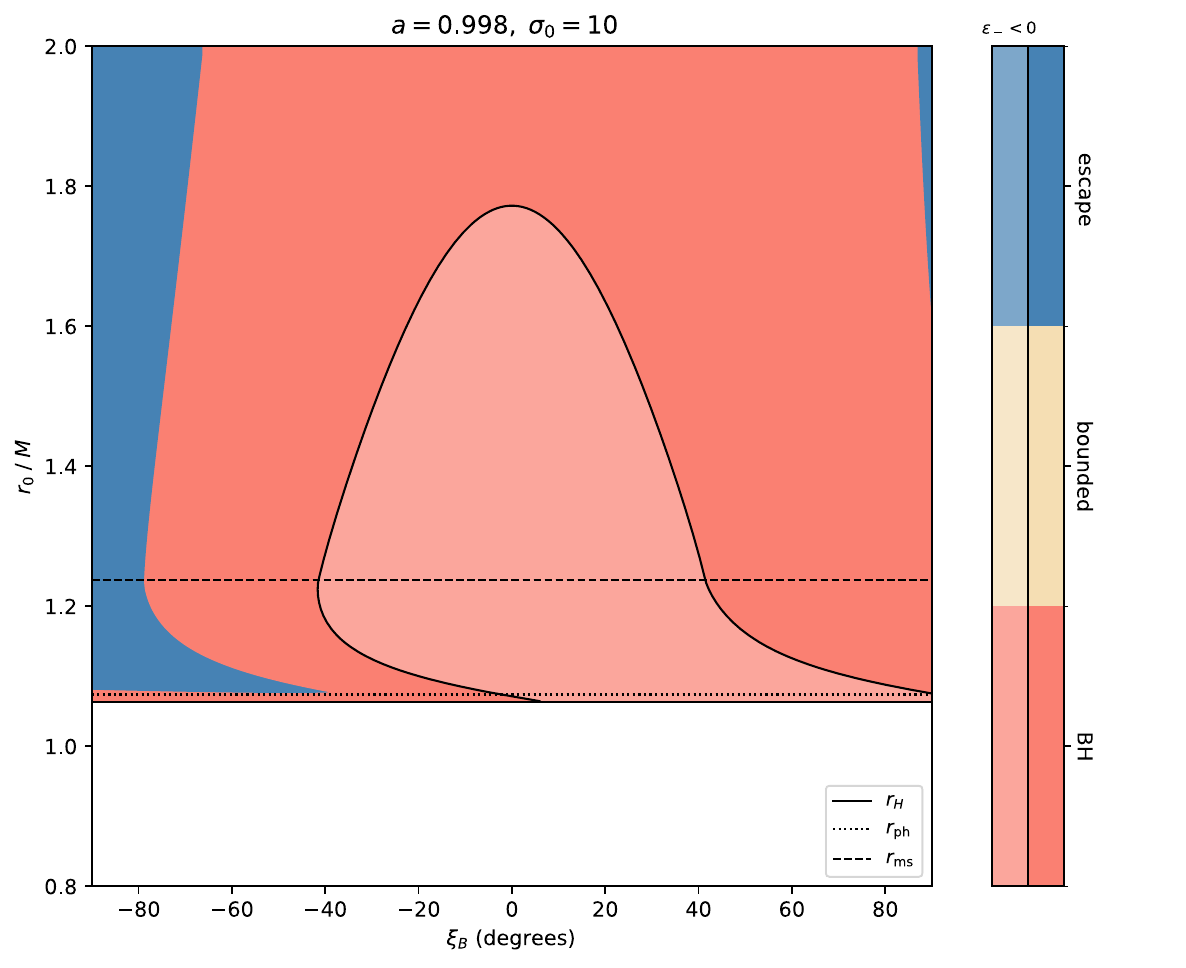}
    \includegraphics[width=0.32\linewidth]{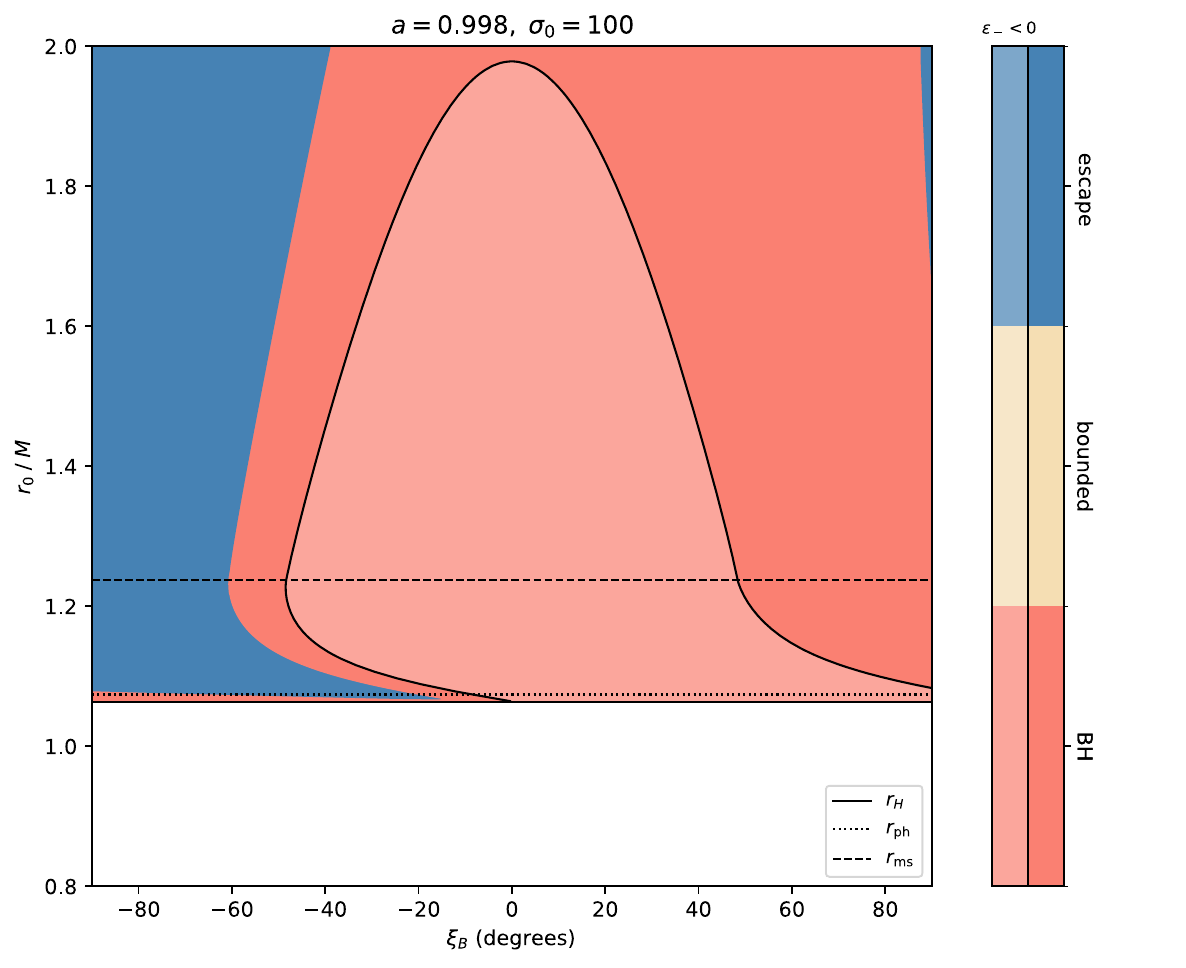}
    \caption{The $\xi_B-r_0$ parameter planes on which the escaping p$^-$, bounded p$^-$ and those that plunge into the central black hole are colored in blue, yellow and red, respectively. The boundaries of energy extraction regions are represented by solid curves. The dotted and dashed straight lines represent the photon sphere and ISCO, respectively. We exhibit the results of $a=0.998$ only. The left, central and right panels exhibit the results of $\sigma_0=$ 2, 10 and 100, respectively.}
    \label{fig:xi-r0-neg}
\end{figure*}

In Fig.~\ref{fig:xi-r0-neg}, we exhibit the destinies of p$^-$ in the $\xi_B-r_0$ parameter planes for $a=0.998$ and $\sigma_0=$2, 10, and 100. We can see that most p$^-$ are absorbed by the central black hole although the centrifugal effect is substantially strong. Moreover, it is confirmed from Fig.~\ref{fig:xi-r0-neg} that plunging to the central black hole is the only destiny of p$^-$ with negative energy-at-infinity density, even in the limit of high magnetization.

\bibliography{Refs}

\end{document}